    \documentclass{aa}  

\usepackage{txfonts}
\usepackage{graphicx,lipsum}
\usepackage[percent]{overpic}
\usepackage{amssymb,amsmath}
\usepackage[version=4]{mhchem}
\usepackage{newtxtext}
\usepackage[varvw]{newtxmath}
\usepackage{xcolor}
\usepackage{caption,subcaption}

\newcommand{\ignore}[1]{}

\def\aap {{A\&A}}

\def\apjl {{ApJL}}

\def\apjs {{ApJS}}

\begin{document}

   \title{ Planet formation in the PDS 70 system }
   \titlerunning{ Constraining the atmospheric chemistry of PDS 70b and c }

    \subtitle{ Constraining the atmospheric chemistry of PDS 70b and c }

   \author{Alex J. Cridland\inst{1,2,3}\thanks{alexander.cridland@oca.eu}, Stefano Facchini\inst{4,6}, Ewine F. van Dishoeck\inst{3,5}, Myriam Benisty\inst{1,2}
          }
    \authorrunning{ Cridland et al. }

   \institute{
   $^1$Université Côte d'Azur, Observatoire de la Côte d'Azur, CNRS, Laboratoire Lagrange, 96 Bd. de l'Obseravtoire, 06300 Nice, France\\
   $^2$Univ. Grenoble Alpes, CNRS, IPAG, 414 Rue de la Piscine, 38000 Grenoble, France\\
   $^3$Max-Planck-Institut f\"ur Extraterrestrishe Physik, Gie{\ss}enbachstrasse 1, 85748 Garching, Germany \\
   $^4$European Southern Observatory, Karl-Schwarzschild-Str. 2, 85748 Garching, Germany\\
   $^5$Leiden Observatory, Leiden University, Niels Bohrweg 2, 2300 RA Leiden, the Netherlands\\
   $^6$Dipartimento di Fisica, Universit\'a degli Studi di Milano, via Celoria 16, Milano, Italy
             }

   \date{Received \today}


  \abstract
  {
Understanding the chemical link between protoplanetary disks and planetary atmospheres is complicated by the fact that the popular targets in the study of disks and planets are widely separated both in space and time. The 5 Myr PDS 70 systems offers a unique opportunity to directly compare the chemistry of a giant planet's atmosphere to the chemistry of its natal disk. To that end, we derive our current best physical and chemical model for the PDS 70 disk through forward modelling of the $^{12}$CO, C$^{18}$O, and C$_2$H emission radial profiles with the thermochemical code DALI and find a volatile C/O ratio above unity in the outer disk. Using what we know of the PDS 70 disk today, we analytically estimate the properties of the disk as it was 4 Myr in the past when we assume that the giant planets started their formation, and compute a chemical model of the disk at that time. We compute the formation of PDS 70b and PDS 70c using the standard core accretion paradigm and account for the accretion of volatile and refractory sources of carbon and oxygen to estimate the resulting atmospheric carbon-to-oxygen number ratio (C/O) for these planets. Our inferred C/O ratio of the gas in the PDS 70 disk indicates that it is marginally carbon rich relative to the stellar C/O = 0.44 which we derive from an empirical relation between stellar metallicity and C/O. Under the assumption that the disk has been carbon rich for most of its lifetime, we find that the planets acquire a super-stellar C/O in their atmospheres. If the carbon-rich disk is a relatively recent phenomenon (i.e. developed after the formation of the planets at $\sim 1$ Myr) then the planets should have close to the stellar C/O in their atmospheres. This work lays the groundwork to better understand the disk in the PDS 70 system as well as the planet formation scenario that produce its planets.
}


    \maketitle
%

\section{Introduction}\label{sec:intro}

The link between the gas and ice chemistry in protoplanetary disks and the resulting chemical structure in the atmospheres of (giant) exoplanets has become an important tool in probing the underlying physics of planet formation. The argument follows that chemical gradients in the disk gas and ice - particularly in the carbon and oxygen tracers - are `encoded' into the planet's atmosphere as it accretes its atmosphere-forming gas. As a result, if accurate measurements of the atmospheric number ratio of carbon and oxygen atoms can be made then one can infer from where in the disk the planet has accreted its material. Restricting from where known planets form helps to constrain planet formation models both in terms of their rates - growth vs. migration - as well as their initial conditions. 

Young planets, with ages less than $\sim 10$ Myr, offer a unique view on the connection between planet formation and the chemistry of planetary atmospheres. First, they are still self-luminous because of their latent heat of formation which makes direct spectroscopy (and imaging) more feasible than Gyr-old planets. Second, and in particularly for the ideal case of PDS 70, their natal protoplanetary disk remains sufficiently massive that it can be detected and chemically characterised. In this way we can directly measure chemical properties of the gas that is feeding the planetary atmospheres and use this knowledge to infer the chemical properties of the gas that had fed the planet during its growth phase. Unfortunately, the presence of the disk makes the first point more complex, since it contaminates attempts at direct spectroscopy (discussed more below). We focus more on the second point in this work. Finally, these young, warm planets have not had sufficiently long for internal processes like mixing and sedimentation to drastically alter the observable chemical properties of its atmospheres. We discuss this point in more detail in Section \ref{sec:discussion} below.

The carbon-to-oxygen number ratio or more commonly the carbon-to-oxygen ratio (C/O), is the total number of carbon atoms divided by the total number of oxygen atoms in either the refractory (rock) and/or volatile (gas and ice) components. It has become the most popular elemental ratio for tracing the formation of planets since its introduction by \cite{Oberg11}. The reasons for its popularity are three-fold: first, carbon and oxygen are the most abundant elements heavier than hydrogen and helium in the Universe and thus their molecular forms H$_2$O, CO, CO$_2$ tend to be the most abundant molecules after H$_2$ in astrophysical settings. Second, these same molecular species have strong molecular features in the near- and mid-infrared that can be observed in spectroscopic studies through either direct observations or during a stellar occultation \citep[transit, see the review by][]{Madhu2019}. Finally, due to their differences in condensation temperature there is the aforementioned C/O ratio gradient as a function of radius through the protoplanetary disk. And as such, different formation models, which predict different starting locations for a giant planet should predict different (observable) chemical differences in the atmospheres of gas giant planets.

Measuring the planet's atmospheric C/O ratio offers a link to the \textit{local} C/O ratio in the protoplanetary disk where the giant planet drew down the bulk of its gas. Given that the later stages of gas accretion are typically very rapid, with timescales that are much less than the migration timescales \citep{Pollack1984,Pollack1996,Crid16a,NGPPS1} the atmospheric C/O ratio is determined by a narrow range of radii in the protoplanetary disk. To zeroth order, the local C/O ratio in the gas is determined by the freeze-out of the most abundant volatiles at their `ice lines'. The most famous ice lines are the water ice line near a gas temperature of $\sim 150$ K and the CO ice line near a temperature of $\sim 20$ K. The former for its importance in setting the water abundance of planetesimals in the Solar system \citep[][and references therein]{CieslaCuzzi2006} while the latter is relevant due to its in-direct detection in protoplanetary disks \citep[for ex. in TW Hya,][]{Qi2013b}. However other ice lines, like the CO$_2$ ice line, will also contribute to the overall evolution of carbon and oxygen bearing species in planet forming regions \citep[see for ex.][]{Eistrup2018}.

The \textit{local} C/O ratio ultimately depends on chemical processes impacting the arrangement of carbon and oxygen atoms between the gas and ice phases as well as the total number of carbon and oxygen atoms in the system. The latter can be represented by a \textit{global} C/O ratio and be thought of as the C/O ratio of the gas and ice of the stellar system's primordial molecular cloud. Given that the disk and star accreted from the same molecular cloud material it is often assumed that the stellar C/O ratio represents this global C/O ratio. Having a measure of the \textit{global} C/O ratio via the carbon and oxygen abundances in the host star is crucial to understanding a planet's atmospheric C/O ratio in the context of its planetary system and formation processes. 

The paper is organised as follows: in section \ref{sec:system} we outline the important features of the PDS 70 system and in section \ref{sec:methods} we outline our numerical and analytic methods. The results are summarised in section \ref{sec:results} and are discussed, along with the model and assumption caveats, in section \ref{sec:discussion}. We concluded our results and offer direction for future study in section \ref{sec:conclusion}.

\section{ The PDS 70 system }\label{sec:system}

\subsection{ The star }

\ignore{ 
PDS 70 was first included in the Cordoba Durchmusterung catalogue \citep{CD1894} from the National Argentine Observatory and given the label of CD-40 8434. Its infrared excess was detected by the Infrared Astronomical Satellite (IRAS) where it was included in volume 7: the Small Scale Structure Catalogue \citep{iras88}, receiving the label IRAS 14050-4109. It was included in the search for T Tauri stars based on IRAS point sources by \cite{GH1992} using the 1.6 m telescope of the Laborat\'orio Nacional de Astrof\'iscia at Pico dos Dias, Brazil. This Pico dos Dias Survey (PDS) identified it as a newly discovered T Tauri star and gave it its commonly used label.
}

PDS 70 was first identified as a T Tauri star in the Pico dos Dias Survey (PDS) by \cite{GH1992}. It was included in the Radial Velocity Experiment (RAVE) fourth \citep{Rave4} and sixth \citep{Rave6} date releases where it was measured at medium resolution ($R\sim$ 7500) to spectroscopically derive its stellar atmospheric properties such as effective temperature, gravitational acceleration and metallicity. It has an effective temperature of $4237 \pm 134$ K, a $\log g = 4.82 \pm 0.2$ and $[{\rm Fe/H}] = -0.11 \pm 0.1$. It is at a distance of 113 pc \citep{GaiaDR32020}.

For the purpose of this paper, it would be useful to have an estimate of the photospheric abundance of carbon and oxygen of the star. Baring this measurement, we can estimate the stellar C/O based on its inferred metallicity and the relation between these values in stars found by \cite{SA2018}. In their sample, they found a linear relation between the star's metallicity and its C/O - particularly between stars that host giant planets (as PDS 70 does). Using their fitted slope ($0.41 \pm 0.02$) and a normalisation of C/O $= 0.48 \pm 0.15$ at $[{\rm Fe/H}] = 0$ - which is slightly lower than the commonly accepted solar value of 0.54 - we find a stellar C/O = $0.44 \pm 0.19$ for PDS 70.

\subsection{ The disk }

The infrared excess identified in the IRAS catalogue by \cite{GH1992} classified PDS 70 as a T Tauri system and implies, given our current knowledge of T Tauri systems, the presence of a dusty disk. The excess was confirmed by \cite{MSH2004} and the disk was soon after detected in coronagraphic images of its scattered light by \cite{Riaud2006}. The inner gap and transition disk-nature of the PDS 70 disk was confirmed through $H$-band polarmetric and $L^\prime$-band imaging by \cite{Hashimoto2012}, and follow up observations at 1.3 mm with the Sub-Millimeter Array (SMA) additionally found evidence for a compact disk inside the dust gap \citep{Hashimoto2015}. The disk was studied at a higher spatial resolution with the Atacama Large (sub)Millimetre Array (ALMA) at 0.87 mm continuum and CO $J=3-2$ and HCO$^+$ $J=4-3$ line emission by \cite{Long2018}. Similar to the dust the gas emission shows a gap near the location of the two planets \citep[][and see below]{Keppler2019}, it also extends much farther ($\sim 200$ AU) from the host star than the dust ($\sim 80$ AU). 

High resolution observations of the system with VLT/SPHERE and VLT/NaCo have mapped the small dust distribution around the host star in more detail and have confirmed the presence of the inner dust disk \citep{Keppler2018}. The size of the dust disk was constrained to less than 17 AU based on the polarized intensity of the dust emission, but could not be further constrained because of a degeneracy between the outer radius and dust depletion factor in their inner disk model. Along with the more detailed look at the physical properties of the PDS 70 disk, \cite{Keppler2018} also report the first detection of an embedded planetary companion to the PDS 70 star. Now called PDS 70b, the direct detection of this embedded planet is a first of its kind. This detection was later confirmed by \cite{Haffert2019} who also detect a second embedded planet, PDS 70c, using H$\alpha$ emission.

\subsection{ The planets }

The detection of PDS 70b \citep{Keppler2018} and PDS 70c \citep{Haffert2019} represents a unique opportunity in the study of planet formation. At time of writing, they are the two only confirmed embedded planets that have been directly detected\footnote{Some recent observations have suggested the detection of point-like feature in AB Aur \citep{Currie2022}, however this is debated \citep{Zhou2022}.} by the emission of their young atmospheres \citep{Wang2021}, through the H$\alpha$ emission of currently accreting gas \citep{Haffert2019}, and through continuum emission coming from the circumplanetary disk \citep{Isella2019,Benisty2021}. Because of their coexistence with their natal protoplanetary disk and the inferred age of the system ($\sim 5$ Myr) they are likely the youngest exoplanets ever detected, in a stage of evolution consistent with the final stages of planet formation. At this final state, the planets have opened a gap in the disk because its gravitational influence has exceeded the viscous and gas pressure force otherwise governing hydrodynamics. The planets have thus largely accreted the bulk of their gas, with any remaining accretion being fed by gas that approaches and then surpasses the gap edge through meridional flows \citep{Morbidelli2014,Teague2019}. We quote the planetary properties of the two embedded planets as derived by \cite{Wang2021} who inferred the orbital parameters based on the assumption that the system is dynamically stable over the lifetime of the system ($\sim 5$ Myr).

\begin{table}[]
    \centering
    \caption{Planetary properties derived by \cite{Wang2021}}
    \begin{tabular}{c|c|c}
    \hline\hline
    & PDS 70b & PDS 70c \\\hline
    Mass (M$_{\rm Jup}$)     & 3.2$^{+8.4}_{-2.1}$ & 7.5$^{+7.0}_{-6.1}$  \\
    Semi-major axis (AU)    & 20.8$^{+1.3}_{-1.1}$ & 34.3$^{+4.6}_{-3.0}$ \\\hline 
    \end{tabular}
    \tablefoot{The values are based on the requirement of a dynamically stable system over the age of the system. The uncertainties are based on their 95\% confidence intervals.}
    \label{tab:the_planets}
\end{table}

Both planets were observed in the near infrared at high spatial and spectral resolution using the GRAVITY instrument on the Very Large Telescope Interferometer (VLTI) by \cite{Wang2021}. Their Spectral Energy Distribution (SED) fitting included the Exo-REM models of \cite{Charnay2018} which allows an inference of the atmospheric C/O of both planets. Unfortunately their `adequate' models - which have Bayes factors within a factor of 100 of the best fit - could not provide stringent restrictions on the atmospheric C/O other than a constraint of C/O$>0.4$ in PDS 70b. For PDS 70c, none of the Exo-REM models were considered `adequate', however the best fit Exo-REM model - with a Bayes factor of 114$\times$ smaller than the best fit model - constrained the atmospheric C/O$<0.7$. This constraint, however, lies very close to the upper limit in their model's C/O prior and thus should be taken with a grain of salt.

\cite{Cugno2021} similarly struggled to place constraints on the chemical properties of the planets in the PDS 70 system. They used medium resolution data from VLT/SINFONI and the `molecular mapping' method \citep{Hoeijmakers2018} to try and find molecular emission from the planets. They predict relatively high upper limits on the molecular abundances of CO and H$_2$O ($10^{-4.1}$ and $10^{-4.0}$ relative to hydrogen respectively) which is in contention with the results of \cite{Wang2021}. A possible reason for deviations between the two observations could be extinction either by clouds or surrounding dust which impacts the overall efficiency of the molecular mapping technique.

\subsection{ Summary }

For the sake of chemical characterisation there is sufficient data to model the global chemical structure in the PDS 70 system. The stellar C/O = 0.44 and the detection of carbon-rich molecular species like C$_2$H implies that the disk has likely undergone some form of chemical evolution to remove gaseous oxygen from the disk's emitting layer. This chemical processing of oxygen - mainly in CO and H$_2$O - out of the gas phase is known to result in unexpectedly low CO emission in protoplanetary disks \citep{Bruderer2012,Favre2013,Du2015,Kama2016b,Ansdell2016,Bosman2018,Schwarz2018,Krijt2020,Bosman2021MAPS,Miotello2022}. We thus interpret the chemical difference between the stellar C/O and the expected high C/O of the disk \citep[as discussed by ][]{Facchini2021} as the chemical processing of gaseous CO in the ice phase and the locking up of carbon and oxygen rich ices into pebbles that settle to the disk midplane. 

These ices are thus invisible to detection via line emission observations but could, in principle, be available for the accretion into the atmosphere of the young planets. Given that CO is also an abundant carrier of carbon, one would expect a similar depletion of carbon as oxygen. As a result many works that explore disk chemistry in situations where C/O $> 1$ find that a depletion of both carbon and oxygen is needed to explain observed fluxes \citep{Bergin2016,Miotello2019,Oberg2021,Bosman2021MAPS}. We find here that such a carbon depletion is not consistent with the disk line emission observations of \cite{Facchini2021} and the models presented here, and will discuss below.

In order to constrain the chemical properties of the disk we use the line emission survey of the PDS 70 system of \cite{Facchini2021} which included the three most abundant CO isotopologues and many bright hydrocarbon lines. In fact, the detection of bright lines from C$_2$H $J=7/2-5/2$, c-C$_3$H$_2$ $J=3_{21}-2_{12}$, and H$^{13}$CN $J=3-2$ is suggestive of high carbon abundance, relative to oxygen, in the outer disk. In this work, we will model the volatile chemistry and line emission for a subset of the detected species - $^{12}$CO, C$^{18}$O, and C$_2$H - which broadly constrain the temperature, density, and chemical properties of the PDS 70 disk respectively.

Our strategy will proceed as follows: we model the line emission observations of \cite{Facchini2021} to estimate the current physical and chemical structure of the PDS 70 disk. We will stay close to the derived physical structure of \cite{Keppler2019} who studied the dust continuum and CO line emission of the disk. We will use the derived physical structure to estimate the physical structure of the disk back when the planets presumably began forming and compute the chemical structure of this younger disk. Finally we will compute the formation of the two planets over the lifetime of the PDS 70 system ($\sim 5$ Myr) to estimate their current-day atmospheric C/O. We will expand on the above steps in more detail below.

\section{ Methods }\label{sec:methods}

This work combines a number of numerical and theoretical work to understand and model the physical and chemical properties of the PDS 70 system. The chemical modelling is done using the Dust And LInes (DALI) code \citep{Bruderer2012,Bruderer2013} which computes the chemical and thermal evolution of the protoplanetary disk gas and ice self-consistently. The DALI code also contains a radiative transfer module for computing the radiative heating and cooling (for the thermal evolution), but is also used to produce synthetic observations of the disk models. 

\subsection{ `Current' disk model }

Our first step is to determine the `current' physical and chemical structure of the PDS 70 disk at its current age of 5 Myr based on the line observations of \cite{Facchini2021}. The gas disk model is based on a simple parametric description of a transition disk proposed by \cite{Andrews2011} and is discussed in detail in \cite{Bruderer2013}. The model follows the standard self-similar solution of a viscously evolving protoplanetary disk assuming that the gas viscosity is constant with time and radially varies as a power-law \citep{LB74}. Solutions of this form follow:\begin{align}
    \Sigma_{\rm gas}(R) = \Sigma_c \left(\frac{R}{R_c}\right)^{-\gamma} \exp{\left\{-\left(\frac{R}{R_c}\right)^{2-\gamma}\right\}},
    \label{eq:main_gas}
\end{align}
where $\Sigma_c$ is the `critical surface density' at a `critical radius' of $R_c$. The capital $R$ denotes the midplane radius. We allow the gas and dust density to extend out to a maximum radius of $R_{\rm disk}$.

Transition disks are characterised by the presence of a large dust gap caused by (at least in the case of PDS 70) the presence of giant planets. This gap can extend all the way to the host star (in which case it is colloquially called a cavity) but transition disks can also contain a smaller inner disk of gas and dust. Thus rather than a radially constant dust-to-gas ratio $\Delta_{\rm dtg}$, the model proposed by \cite{Andrews2011} and used in DALI reduces $\Delta_{\rm dtg}$ to arbitrarily low values between the outer radius of the inner disk ($R_{\rm gap}$) and the inner radius of the outer disk disk ($R_{\rm cav}$).

Inward of $R_{\rm cav}$ the gas and dust can also be further depleted relative to their expected density from Equation \ref{eq:main_gas}. For the gas this is justified by the assumption that gas accretion is slightly less efficient inside the dust cavity due to the gravitational influence of the embedded planets, hence $\Sigma_{\rm gas}^\prime(R < R_{\rm cav}) = \delta_{\rm gas}\Sigma_{\rm gas}(R)$. The dust density has the form:\begin{align}
    \Sigma_{\rm dust} = \begin{cases}
            \delta_{\rm dust}\Delta_{\rm dtg}\Sigma_{\rm gas}(R) & R < R_{\rm gap}\\
            10^{-15} & R_{\rm gap} < R < R_{\rm cav}\\
            \Delta_{\rm dtg}\Sigma_{\rm gas}(R) & R > R_{\rm cav}
        \end{cases}.
    \label{eq:dust_cases}
\end{align}
The final two important radii for the model are the sublimation radius $R_{\rm subl} \sim 0.07\sqrt{L_*/L_\odot}$ AU inward of which the disk is sufficiently warm for the dust to evaporate \citep{Bruderer2013} and the outer radius $R_{\rm out}$ which describes the largest radius in the simulation. For the stellar spectrum we use a blackbody with effective temperature noted above. The star has a bolometric luminosity of 0.35 L$_{\odot}$.

The vertical distribution of the gas follows a Gaussian profile with a physical scale height following $H = H_c (R/R_c)^\phi$. The so-called scale height angle $h\equiv H/R$ is thus $h = h_c (R/R_c)^{\phi-1}$, where the subscript `c' continues to denote the value of any variable at the critical radius $R_c$. The vertical distribution of the dust is more complex than the gas and is split into two populations. The `small' population represents grains with sizes ranging between 0.005 - 1 $\mu$m while the `large' population represents grains 1 - 1000 $\mu$m. While the small grains tend to be well coupled to the gas and thus follow the vertical distribution more closely, the larger grains will tend to settle to the midplane. The vertical volume density of the dust follows, in cylindrical coordinates:\begin{align}
    \rho_{\rm dust,small} &= \frac{(1-f)\Sigma_{\rm dust}}{\sqrt{2\pi}R h} \exp{\left[-\frac{1}{2}\left(\frac{0.5\pi - \theta}{h}\right)^2\right]} ~{\rm and}\nonumber\\
    \rho_{\rm dust,large} &= \frac{f\Sigma_{\rm dust}}{\sqrt{2\pi}R\chi h}\exp{\left[-\frac{1}{2}\left(\frac{0.5\pi - \theta}{\chi h}\right)^2\right]},
    \label{eq:vert_dust}
\end{align}
where $\theta = \arctan{(R/z)}$, $\chi< 1$ parameterises the amount of settling that occurs for the large dust grains and $f$ sets their mass fraction. DALI will run in two modes, the first takes Equations \ref{eq:main_gas}-\ref{eq:vert_dust} with a list of prescribed parameters to produce the disk gas and dust distributions while the second allows users to prescribe their own choice of dust and gas radial and vertical distribution (discussed more below). We show the adopted values for the discussed parameters in Table \ref{tab:disk_params}. In the second column of the table, if a single value is shown it coincides with the derived disk model of \cite{Keppler2018} based on the dust distribution. 

\begin{table}
    \centering
    \caption{Range of disk parameters used to construct current disk model.}
    \begin{tabular}{c|c}
    \hline\hline
        Parameter & Range of values \\\hline
        $\Sigma_{c}$ & 2.87 g cm$^{-3}$ \\
        $R_c$ & 40 AU \\
        $f$ & 0.968 \\
        $\chi$ & 0.2 \\
        $\gamma$ & 1 \\
        $\phi$ & 0.25 \\
        $R_{\rm gap}$ & 2.0, 4.0, \textbf{6.0}, 8.0, 10.0 AU \\
        $R_{\rm cav}$ & 45 AU \\
        $R_{\rm disk}$ & 120 AU \\
        $\delta_{\rm gas}$ & 0.000001,0.0001,\textbf{0.01}, 1.0 \\
        $\delta_{\rm dust}$ & 0.000001,\textbf{0.0001}, 0.01, 1.0 \\
        $\delta_{\rm planet}$ & 1.0, $\bf{10^{-2.5}}$, $10^{-15}$ \\
    \hline
    \end{tabular}
    \tablefoot{Important parameters for the construction of the transition disk model of PDS 70, including the range of values used to determine the current disk model. The parameters without ranges in the second column coincide with the derived properties of the disk by \cite{Keppler2018}. Bolded values show the parameters of our preferred model.}
    \label{tab:disk_params}
\end{table}

The quantity of gas within the orbital range of the planets and the dust cavity is still an open question. Numerical simulations \citep[eg. ][]{Lubow2006} have shown that gas accretion can occur across the planet-induced gap, but results in a lower accretion rate onto the host star than the accretion rate outward of the gap. Theoretical studies have shown \citep[for ex. see ][]{Crida2006} that a giant planet modifies the local gas density surrounding it as its gravitational torque competes against the viscous torques of the surrounding protoplanetary disk. It is thus expected that the gas surface density should be lower around the embedded planets of the PDS 70 system. 

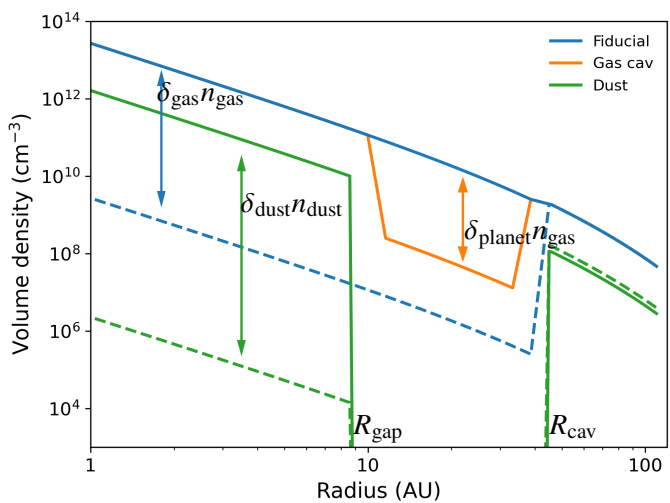
\begin{figure}
    \centering
    \begin{overpic}[width=0.5\textwidth]{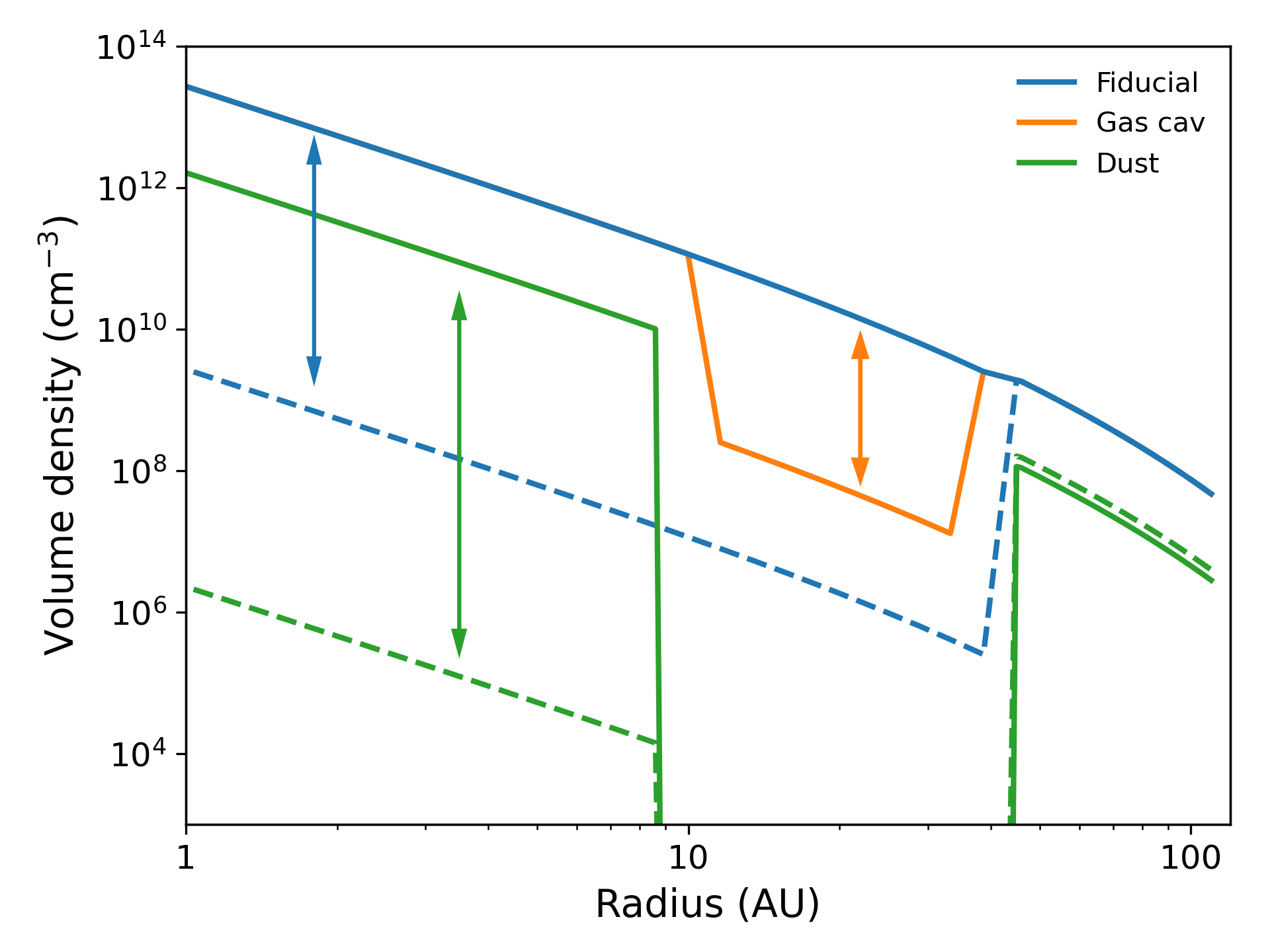}
    \put(24,60){\large $\delta_{\rm gas}n_{\rm gas}$}
    \put(36,44){\large $\delta_{\rm dust}n_{\rm dust}$}
    \put(68,40){\large $\delta_{\rm planet}n_{\rm gas}$}
    \put(52,12){\large $R_{\rm gap}$}
    \put(80,12){\large $R_{\rm cav}$}
    \end{overpic}
    \caption{Examples of possible models of the gas and dust distributions to be fed into DALI. Shown is the number volume density defined as n$_{\rm gas} = \Sigma_{\rm gas}/(\sqrt{2\pi}H\mu m_{\rm H})$ for the gas. For the dust, $n_{\rm dust}$ is defined to ensure the correct dust-to-gas mass ratio. The solid and dashed curves show the effect of depleting the inner disk by constant factors of $\delta_{\rm gas}$ and $\delta_{\rm dust}$ while the orange curve show the impact of adding a planet-induced gas gap of a prescribed depth of $\delta_{\rm planet}$.}
    \label{fig:cavity_test}
\end{figure}

\cite{Keppler2019} used $^{12}$CO $J=3-2$ line emission data to constrain the gas properties within the dust cavity. They found evidence of a gas gap around the PDS 70b planet consistent with a mass of 5 $M_{\rm Jup}$. No gas gap was inferred for PDS 70c, however the numerical simulations of \cite{Bae2019} showed that the two planets would share the gas gap. In this work we can include the impact on the line emission of CO and C$_2$H from the presence of a planet-driven gap. In order to add the impact of the gas gap we further deplete the gas density within the dust cavity by a factor of $\delta_{\rm planet}$ (see Figure \ref{fig:cavity_test}). 

In Figure \ref{fig:cavity_test} we show an example of how the different forms of depletion impact the structure of the midplane volume (number) densities. While the blue and green solid lines represent the fiducial distribution of the gas and dust (respectively) in a transition disk model, their dashed lines show the effect of reducing the inner disk densities by $\delta_{\rm gas} = \delta_{\rm dust} = 0.01$. Meanwhile the orange line shows the impact of imposing a reduction in the gas density by a factor of $\delta_{\rm planet} = 10^{-2.5}$ within the dust cavity, which approximately equates to the 5 M$_{\rm Jup}$ model of \cite{Keppler2019}.

\begin{figure}
    \centering
    \begin{overpic}[width=0.5\textwidth]{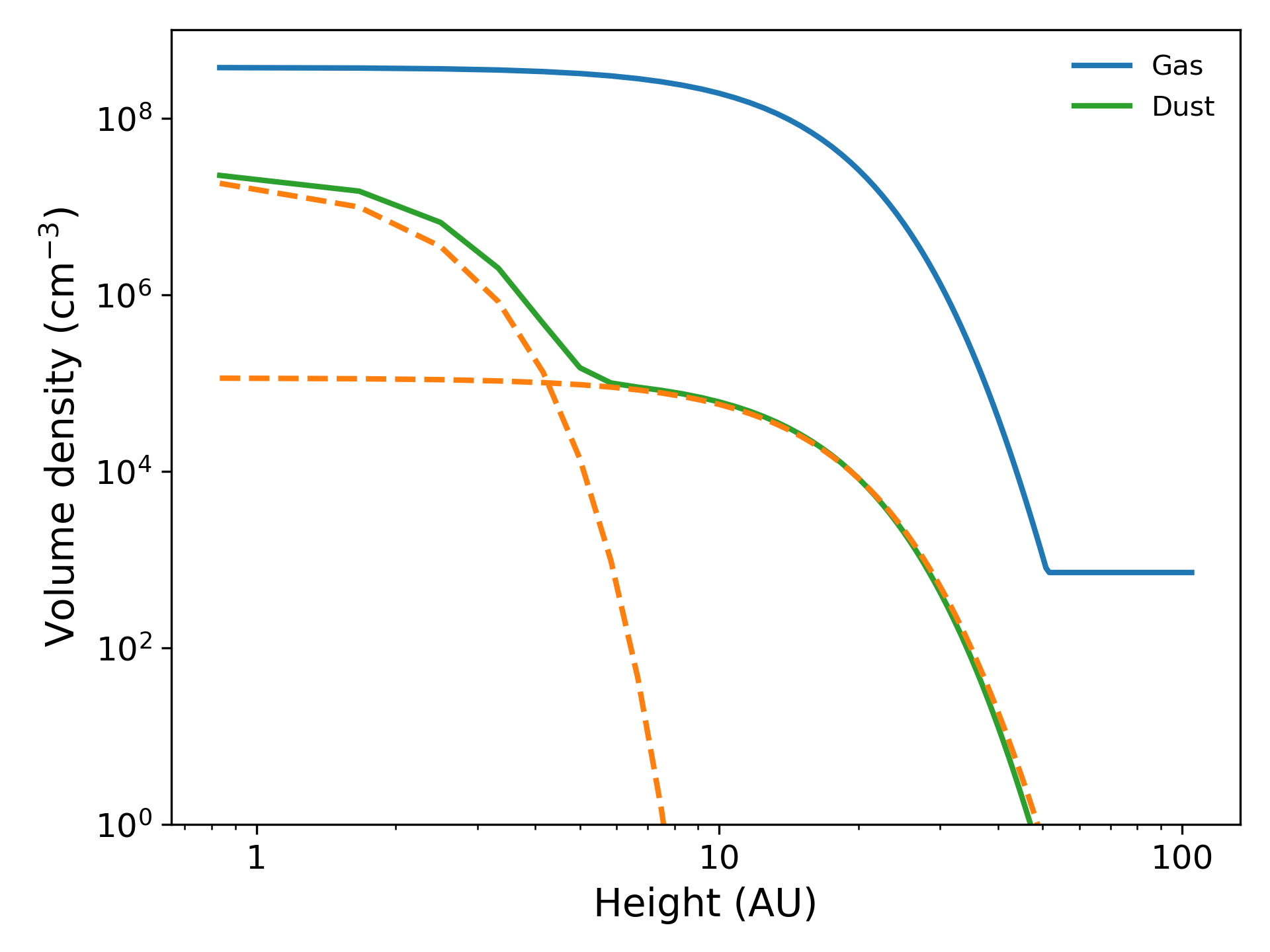}
    \put(23,50){\large $n_{\rm dust,large}$}
    \put(59,25){\large $n_{\rm dust,small}$}
    \end{overpic}
    \caption{The vertical distribution of the gas and dust at a radius of $\sim 80$ AU - outward of the dust cavity outer edge. The orange dashed line represent the different distributions of the large and small dust (eq. \ref{eq:vert_dust}). The large dust grains are distributed closer to the midplane because they tend to settle with respect to the smaller grains. As a result the dust-to-gas ratio is slightly higher near the midplane when compared to far above the midplane. }
    \label{fig:vert_test}
\end{figure}

In Figure \ref{fig:vert_test} we show an example of the vertical distribution of the gas and dust in a fiducial model at a radius of $\sim$ 80 AU, outward of the cavity outer edge. The gas vertical distribution (blue curve) is described by a single Gaussian profile while the dust vertical profile (green line) is described by a pair of Gaussian profiles (ie. Equation \ref{eq:vert_dust}). The orange dashed lines show each of the Gaussian profiles from Equation \ref{eq:vert_dust}, converted to number density rather than mass density, to demonstrate the fact that the dust profile is effectively the sum of the two Gaussian profiles.

\subsection{ Disk chemistry }

As previously mentioned the chemistry is handled using the DALI \citep{Bruderer2012,Bruderer2013} code. DALI simultaneously computes the gas and dust temperature via radiative transfer (both continuum and line) as well as the chemical evolution of the volatiles molecular species. As a result the code computes an accurate picture of the temperature structure and molecular abundance in the upper atmosphere of the disk while simultaneously predicting the relative molecular line strengths and line emission from a large range of species.

\begin{figure}
    \centering
    \includegraphics[width=0.5\textwidth]{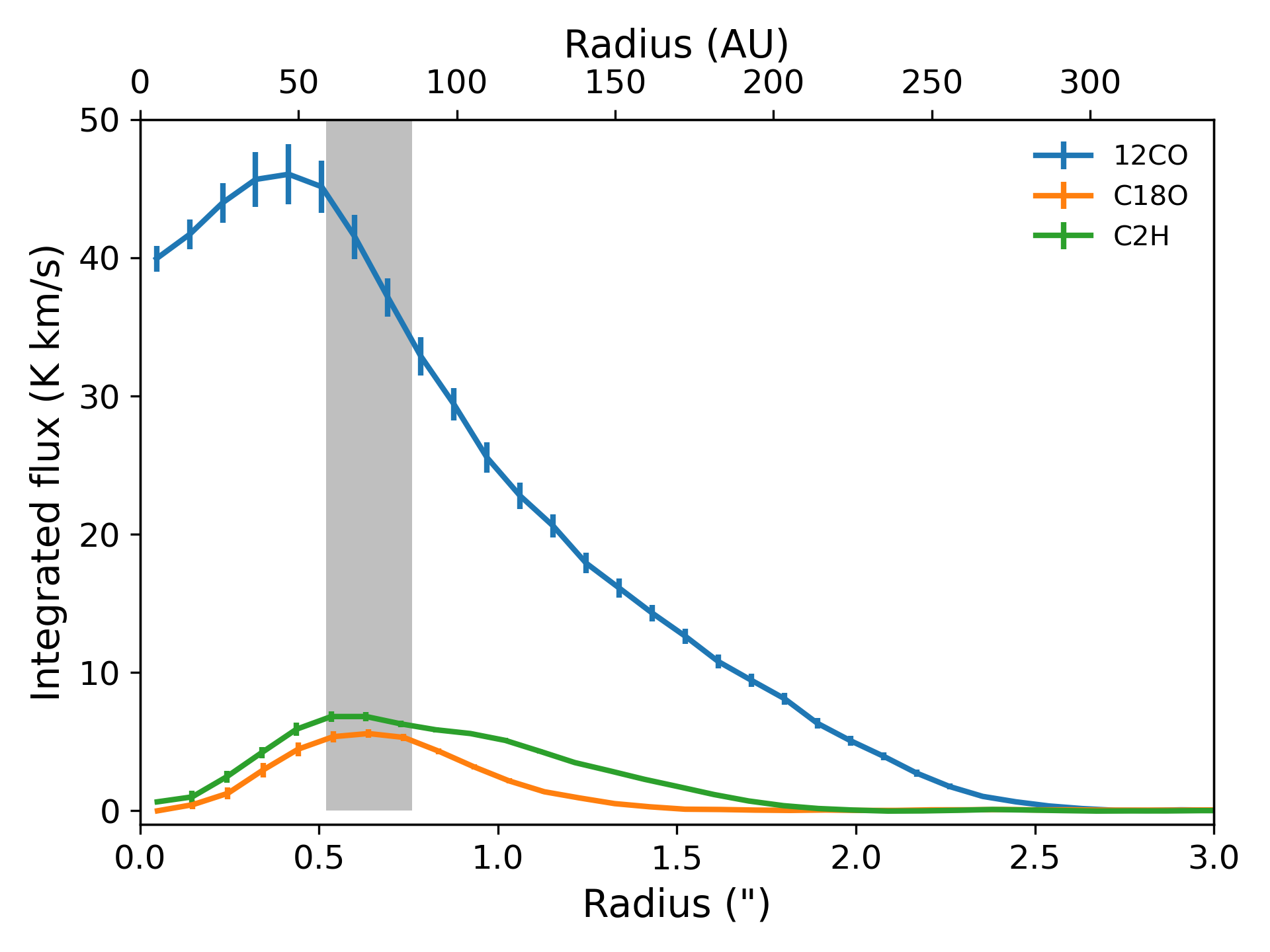}
    \caption{The radial profile observed by \cite{Facchini2021} using ALMA. The grey band shows the approximate location of the (mm) dust ring according to \cite{Keppler2019}. The line fluxes peak in slightly different positions with the optically thin tracers peaking near the edge of the dust cavity and the optically thick tracers peaking inside the cavity.}
    \label{fig:observed}
\end{figure}

In particular we focus on three molecular lines: $^{12}$CO $J=2-1$, C$^{18}$O $J=2-1$, and C$_2$H $J=7/2-5/2$. These molecular lines are sensitive (mainly) to the temperature, density structure, and the chemistry of the disk respectively. These lines, among others, were observed by \cite{Facchini2021} using the Atacama Large (sub)Millimeter Array (ALMA) at a resolution of $\sim 0.4"$, and are shown in Figure \ref{fig:observed}. One of the main features found in this observational work is that at this resolution the integrated intensity is not centrally peaked, instead peaking within the dust cavity. In addition the strong C$_2$H emission is indicative of a large C/O - likely above unity - in order for C$_2$H to have a significant abundance in the disk.

\ignore{
In this work we use the lower resolution ALMA data of \cite{Facchini2021} over the soon-to-be-published high resolution ALMA data of the same target because our goal centres on understanding the global chemical picture in the disk. This global picture is more useful in constraining the chemistry of the embedded planetary atmospheres because it better informs our planet formation models of what the chemical picture of the disk \textit{was} when the planets began to form. With that said, this global picture is less well equipped in modelling the current day chemical properties of the disk because our chemical models ignore any radial transport of volatiles \citep[as seen in for ex.][]{Booth2017,Bosman2017b,Krijt2020}.
}

We focus on constraining the global C/O ratio in the disk as a first step in understanding the chemical environment in which the planets formed. Our chemical model, discussed below, is sensitive to chemistry-driven local changes to the volatile (gas/ice) C/O ratio such as the freeze-out; however we neglect local changes to the C/O ratio that may be driven by the radial transport of volatiles \citep[as seen in for ex.][]{Booth2017,Bosman2017b,Krijt2020}. Given the presence of the two planets, the strong dust trap outside of their orbit, and the evidence of tenuous gas in the inner disk \citep{Keppler2019}, an exploration of the radial dependence of C/O would be an interesting focus for future work.

We employ the chemical network used by \cite{Miotello2019} to model the global C$_2$H emission from an ALMA survey of protoplanetary disks. The chemical network is based on the network originally developed for DALI \citep{Bruderer2012} and pulls reactions and rates from the UMIST Database for Astrochemistry \cite[UDfA, ][]{Wood07,McE03}. It is optimised for the abundances of CO and related species in the gas phase and includes UV cross-sections and photodissociation rates of \cite{Heays2017}. \cite{Visser2018} added to the network to include nitrogen chemistry which often acts to compete with hydrocarbons (like C$_2$H) for available carbon.

While chemical networks that include isotope-specific reactions exist \citep[e.g. ][]{Miotello2016}, we do not include these reactions here. Instead we assume that species like C$^{18}$O scale with their more abundant isotopologues by the standard isotopic ratios in the Interstellar medium (ISM). The isotopic ratios of carbon and oxygen in the ISM that are relevant to our work are [$^{12}$C]/[$^{13}$C] $=77$ and [$^{16}$O]/[$^{18}$O] $=560$ \citep{WR1994}. By simply scaling the abundance of species like C$^{18}$O to the abundance of the standard isotope ($^{12}$C$^{16}$O) we ignore possible chemical effects like isotopic selective photodissociation which causes the less abundant isotopologues to be further depleted because of their less efficient self-shielding. As a result the less abundant isotopologue becomes optically thick lower in the disk atmosphere than would be assumed from simple scaling models. \cite{Miotello2016} showed that this will slightly weaken the line intensity of these less abundant isotopologues because the emission is coming from deeper in the disk atmosphere, from a slightly cooler region of the disk. Furthermore the excess photodissociation results in more free atomic oxygen isotopes that can be incorporated into other species.

We begin the chemical simulations with the majority of chemical elements in the atomic phase \citep[i.e. the \textit{reset} scenario of ][]{Eistrup2016}, except for hydrogen which begins in its molecular form and a fraction of the nitrogen which begins as frozen NH$_3$ and gaseous molecular N$_2$. The initial abundances follow those selected by \cite{Miotello2019} except in the case of carbon and oxygen which is varied. The initial abundances are summarised in Table \ref{tab:init_abun}. Our fiducial model uses the carbon and oxygen abundances shown in the table. These values result in a marginally carbon-rich disk (C/O = 1.01) which is what is expected from the presence of C$_2$H emission. We will also explore a range of C/H and O/H (and thus C/O) in order to confirm this expectation.

\begin{table}[]
    \centering
    \caption{Initial abundances (per number of Hydrogen atoms) used in the fiducial DALI chemical calculation. }
    \begin{tabular}{c|c|c|c}
        \hline\hline
        Species & Abundance & Species & Abundance\\\hline
        H$_2$ & 0.5 & & \\
        He & 0.14 & $J$NH$_3$ & 9.90$\times 10^{-7}$\\
        C & 1.01$\times 10^{-4}$ & H$_3^+$ & 1.00$\times 10^{-8}$\\
        O & 1.00$\times 10^{-4}$ & HCO$^+$ & 9.00$\times 10^{-9}$ \\
        N & 5.10$\times 10^{-6}$ & C$^+$ & 1.00$\times 10^{-9}$\\
        Mg & 4.17$\times 10^{-7}$ & Mg$^+$ & 1.00$\times 10^{-11}$\\
        Si & 7.94$\times 10^{-7}$ & Si$^+$ & 1.00$\times 10^{-11}$\\
        S & 1.91$\times 10^{-6}$ & S$^+$ & 1.00$\times 10^{-11}$\\
        Fe & 4.27$\times 10^{-7}$ & Fe$^+$ & 1.00$\times 10^{-11}$\\
        N$_2$ & 1.00$\times 10^{-6}$ & PAH & 6.00$\times 10^{-7}$\\
        \hline
    \end{tabular}
    \tablefoot{The ions and PAH initialise the total charge and act as a sink for free electrons, respectively. A species with a $J$ label denotes the ice phase of the given species.}
    \label{tab:init_abun}
\end{table}

\subsection{ Disk model at the time of planet formation }\label{sec:dmodel}

More than likely, the physical properties of the PDS 70 disk were different when the planets b and c started forming $\sim 4-5$ Myr ago. The particular `starting time' for planet formation is a remarkably difficult feature to constrain - particularly for exoplanet systems where we lack the isotopic data that is used in Solar system studies on the subject. Here we will assume that the core of both planets formed within the first 1 Myr of the protoplanetary disk lifetime, and begin our gas accretion simulation from there. 

We use analytic models of protoplanetary disk evolution to determine the density structure of the gas at a disk age of 1 Myr. These models, derived by \cite{Tabone2021}, assume that the disk evolves through either viscous evolution or magnetohydrodynamic (MHD) disk winds. It is currently unclear which mechanism is the main driver of angular momentum transport, and thus protoplanetary disk evolution, so modelling both separately accesses the two extremes of disk evolution. The main physical difference between the two mechanisms is the `direction' of their angular momentum transport. Viscous evolution transports the bulk of the angular momentum radially outward along with a small amount of the disk mass, while the bulk of the disk mass moves inward. In other words viscous disks evolve by spreading outward to larger radii with a small fraction of its mass, while the remainder moves inward toward the host star.

Protoplanetary disks evolving due to MHD disk winds move the bulk of their angular momentum to larger heights, by launching gas along the magnetic field lines emerging from the top of the disk. Unlike in viscously evolving disks, wind-driven disks do not spread their mass outward and are thus expected to remain at roughly the same size throughout their evolution \citep{Tabone2021}. For each of the above cases, \cite{Tabone2021} found that the gas surface density and disk size should evolve as:\begin{align}
    \Sigma_{\rm gas} &= \Sigma_c(t)\left(\frac{R}{R_c(t)}\right)^{-1+\xi}\exp{\left(-\frac{R}{R_c(t)}\right)}~ {\rm where}\nonumber\\
    R_c(t) &= R_c(t=0)\left(1 + \frac{t}{(1+\psi)t_{\rm acc,0}}\right)~ {\rm and}\nonumber\\
    \Sigma_c(t) &= \Sigma_c(t=0)\left(1 + \frac{t}{(1+\psi)t_{\rm acc,0}}\right)^{-(\frac{5}{2}+\xi+\frac{\psi}{2})}.
    \label{eq:tabone}
\end{align}
The relevant timescale, $t_{\rm acc,0}$, in these equations is the initial accretion timescale and denotes the time that would require a gas parcel to accrete from an initial position of $R_c/2$ to the inner region of the disk. It has the functional form of:\begin{align}
    t_{\rm acc,0} = \frac{R_c(t=0)}{3h_cc_{s,c}\Tilde{\alpha}(t=0)},
    \label{eq:timescale}
\end{align}
where $c_{s,c}$ is the sound speed at the (initial) critical radius and $\Tilde{\alpha}$ is the effective disk-$\alpha$ \citep{SS73} due to either viscous or MHD disk winds. The parameter $\psi = \alpha_{\rm DW} / \alpha_{\rm SS}$ is the ratio of the effective $\alpha$ due to disk winds (DW) and viscosity \citep[SS, the standard $\alpha$ from][]{SS73}. In the limit of a purely viscous disk $\psi = 0$ while in the opposite limit $\psi = \infty$. In the limit that $\psi \rightarrow \infty$ the time evolution of $R_c$ and $\Sigma_c$ in eq. \ref{eq:tabone} becomes:\begin{align}
    R_c(t) &= R_c(t=0)~ {\rm and}\nonumber\\
    \Sigma_c(t) &= \Sigma_c(t=0)\exp{\left(-\frac{t}{2t_{\rm acc,0}}\right)}.
    \label{eq:wind_only}
\end{align}
The final parameter, $\xi$, denotes the mass ejection index and quantifies the effect on the mass surface density profile from the ejection of mass through the disk wind. The change in surface density profile follows from the significant fraction of the gas that can be lost to the disk wind as it extracts angular momentum. For the case of a pure wind ($\psi = \infty$) it has the form $\xi = 1/[2(\lambda-1)]$ where $\lambda$ is magnetic lever arm parameter and has a typical value of between $2-5$.

\begin{table}[]
    \centering
    \caption{Parameters used in setting the disk properties at a disk age of 1 Myr.}
    \begin{tabular}{c|c|c}
    \hline\hline
    Parameter & Viscous ($\psi = 0$) & Wind ($\psi=\infty$)  \\\hline
        $\Sigma_c$ ($t = 1$ Myr) & 44.7 g cm$^{-2}$ & 21.2 g cm$^{-2}$ \\
        $R_c$ ($t = 1$ Myr) & 13.3 AU & 40.0 AU \\
        $\xi$ & 0 & 0.25 ($\lambda = 3$) \\
        M$_{\rm disk, 1 Myr}$ (M$_\mathcal{J}$)$^1$ & 5.8 & 19.9\\
    \hline
    \end{tabular}
    \tablefoot{$^1$Jupiter masses within 120 AU}
    \label{tab:turnback}
\end{table}

\subsection{ Planet formation }

We use a standard approach to the growth and migration of our planetary embryo. In both disk models presented in Section \ref{sec:dmodel} we initiate planetary embryos at a range of initial radii between 20-80 AU in steps of 2 AU. For completeness we also compute the growth and evolution of planets originating from 1-20 AU in steps of 1 AU. The initial mass of the embryos are set by either the pebble isolation mass - using the prescription of \cite{Bitsch2018} - or the total dust mass exterior to the embryos' initial orbits, which ever is lower.

The gas accretion and planetary migration follows the work of \cite{Cridland2020}, and we summarise the important features of the model in Appendix \ref{sec:PFdetails}. For the purpose of this work, the `chemical accounting' of the carbon and oxygen-bearing species is important to determine the atmospheric C/O. We trace the quantity of mass flux onto the atmosphere through Equations \ref{eq:KH_time}, \ref{eq:Bondi_time}, and/or \ref{eq:MTG}, as well as the rate that the growing planets move through their disk using Equation \ref{eq:type2}. We follow the method of \cite{Cridland2020} to compute the acquisition of carbon and oxygen into the atmospheres, we summarise the relevant information below.

\subsection{ Chemical acquisition }

During the age of planet formation the disk physical and chemical properties are kept constant to make the problem more easily tractable. As was done in \cite{Cridland2020}, such simplicity ignores the impact that an evolving disk can have on the overall evolution of the gas chemistry, but allows us to include the chemical impact of meridional flow on the chemistry of planetary atmospheres. We view this as an appropriate trade-off for this study since ALMA line emission observations tend to probe the chemistry of the disk from where meridional flow originates.

Before a gap in the disk is opened by the growing planet we compute the average carbon and oxygen abundance within the planet's feeding zone - which we equate to the proto-planet's Hill sphere. Once the the gap is open we assume that the gas flow towards the planet follows a meridional flow \citep{Mordasini14,Teague2019} and we assume that the planet-feeding gas comes from between one and three gas scale heights. In this case we average the carbon and oxygen abundance in the gas and ice at the edge of the gas gap over one and three gas scale heights.

We assume any ice frozen onto the population of small dust grains is entrained with the gas and is delivered into the planetary atmospheres. This process occurs both while the planet is embedded in the disk as well as after the gap is opened. The large grain population is not as entrained as the small population, but is nevertheless expected to flow with the gas near the planet. Before the gas gap is opened we assume that ices frozen on the large dust grain population can accrete into the planetary atmosphere at a rate that is reduced relative to the rate of gas accretion by a factor of $1/(1 + St^2)$, where $St$ is the average Stokes number of the population of large dust grains.

The refractory component of the disk is known to include a significant fraction of the overall carbon \citep[around 50\%,][]{Berg15}. During the growth of the atmosphere, we do not include any excess carbon from the dust. The small grains dominate the dust mass above the midplane, in a region that is likely to have been photochemically processed to remove this excess carbon \citep{Anderson2017,Klarmann2018,BinkertBirnstiel2023}. This could have contributed to the high C/O that is observed in the PDS 70 gas disk. The dust near the midplane is dominated by larger grains which could maintain their carbon. While we allow for their accretion, they do not significantly contribute to the carbon abundance of the planetary atmospheres.

\section{Results}\label{sec:results}

\begin{figure*}[h!]
    \centering
    \includegraphics[width=\textwidth]{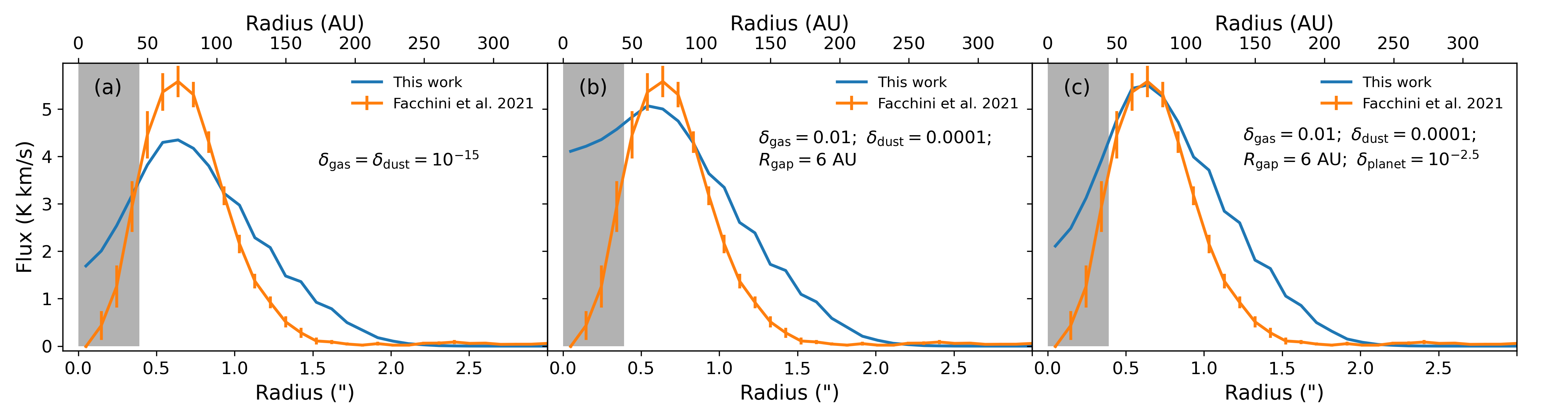}
    \caption{Radial profiles of C$^{18}$O emission for three different disk models compared to the observational data of \cite{Facchini2021}. The grey band shows the width of the major axis of the beam. In each case the relevant parameters are listed in the top right corner of the figure. From left to right: in model (a) the inner disk is almost completely removed, and there is effectively no gas or dust from $R_{\rm cav}$ inward. In model (b) the inner disk is replaced, with the noted depletions of gas and dust. Finally in model (c) the same inner disk parameters are used, however between $R_{\rm gap}$ and $R_{\rm cav}$ the gas is depleted by a further factor. }
    \label{fig:triple_plot}
\end{figure*}

\subsection{Current disk model}

Our first goal is to constrain the chemical and physical properties within the PDS 70 disk by comparing DALI forward models to the observed radial profile of the $^{12}$CO, C$^{18}$O, and C$_2$H emission of \cite{Facchini2021}. We begin with our fiducial model using the parameters derived in \cite{Keppler2019} and neglecting any deviations from the standard surface density profile due to depletion in the inner disk or the embedded planets ($\delta_{\rm gas} = \delta_{\rm dust} = \delta_{\rm planet} = 1.0$). In the following we begin by constraining the density structure using the optically thin C$^{18}$O line. With the constrained density structure we then investigate the impact on the optically thick $^{12}$CO line and the chemical dependence of the optically thin C$_2$H line.

The synthetic images are produced using the radiative transfer package in DALI using the observational parameters of: inclination = 52 degrees, distance = 113 pc. The data are then convolved with the same beam as in \cite{Facchini2021}, de-projected, and averaged over annuli with the same sizes as was done in \cite{Facchini2021}. In addition we use the same channel width and number of channels as is done in \cite{Facchini2021}.

\subsubsection{C$^{18}$O}

In Figure \ref{fig:triple_plot}(a) we show the radial profile of C$^{18}$O emission in the case of a disk that is nearly void of material inward of the outer dust ring edge, while using the remaining fiducial parameters presented in Table \ref{tab:disk_params}. In this case we are exclusively sensitive to emission from the inner edge of the dust ring, and the outer disk. With fiducial parameters $R_{\rm cav} = 45$ AU and $R_{\rm disk} = 120$ AU we find a good agreement between the peak location of the C$^{18}$O emission in the observations (orange line) and our model (blue line). Meanwhile we overestimate the flux coming from the outer disk suggesting that the outer disk is too extended or too diffuse. A smaller $R_{\rm disk}$ may better replicated the observed flux, however since we neglect isotopologue-selected photodissociation, the outer disk could be less abundant in C$^{18}$O than is represented in our model \citep[see for ex.][]{Miotello2014}.

In Figure \ref{fig:triple_plot}(b) we show the C$^{18}$O radial profile for the preferred model which includes an inner disk. Adding a small inner disk into the model adds a new source of optical depth which both cool the inner edge of the dust cavity as well as protect molecular species from photodissociation. Of course the inner disk may also increases the flux of C$^{18}$O at small separations which requires that it is fairly tenuous. The presented model involves a gas depletion $\delta_{\rm gas} = 0.01$ and dust depletion $\delta_{\rm dust} = 0.0001$. The inner disk has a radius of 6 AU in this model.

As a final modification we add an additional gas depletion within the planetary gap caused by the presence of the young pair of giant planets. In Figure \ref{fig:triple_plot}(c) we add an additional depletion $\delta_{\rm planet} = 10^{-2.5}$ into the dust cavity ($R_{\rm gap} < R < R_{\rm cav}$). Reducing the gas density inside the gap leads to a slightly warmer dust wall and a slight increase in the peak flux of C$^{18}$O. In addition the flux coming from the inner regions is lowered back to a similar level as in the model shown in Figure \ref{fig:triple_plot}(a). Because it reproduces the peak flux as well as its location, the model in Figure \ref{fig:triple_plot}(c) is our preferred model.

\subsubsection{$^{12}$CO}

\begin{figure}
    \centering
    \includegraphics[width=0.5\textwidth]{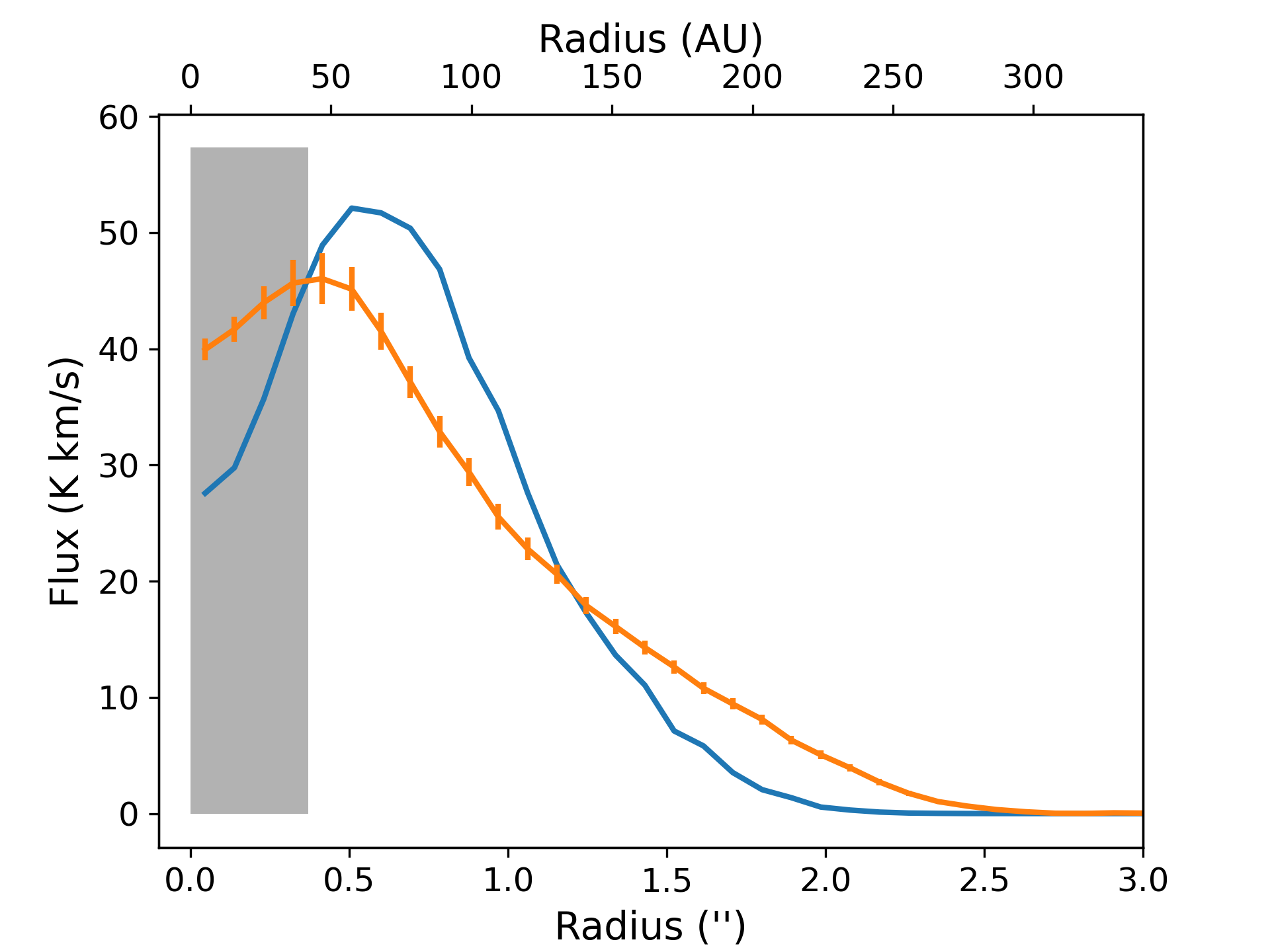}
    \caption{The radial profile of the $^{12}$CO emission due to the model presented in Figure \ref{fig:triple_plot}(c)}
    \label{fig:preferred_plntgap_12CO}
\end{figure}
In Figure \ref{fig:preferred_plntgap_12CO} we show the resulting $^{12}$CO emission radial profile from the disk model presented in Figure \ref{fig:triple_plot}(c). Surprisingly the peak emission is somewhat shifted with respect to the observed radial profile. This may be linked to the exact gas structure near the inner edge of the outer disk. In our model we assume that the gas density sharply cuts off at the inner edge of the outer disk ($R_{\rm cav}$). However its likely that the disk smoothly transitions from the density in the outer disk down to the lower density in the cavity and lower disk \citep{vanderMarel2016,Leemker2022}. The extra width due to this smooth transition could easily contribute to the (optically thick) $^{12}$CO emission while only weakly contributing to the optically thin tracers.

The outer disk flux of $^{12}$CO also falls off more quickly than is seen in observations, which could be exaggerated if the outer disk is truncated as was discussed for C$^{18}$O. Our model over estimates the $^{12}$CO flux between $R_{\rm cav}$ and $R_{\rm disk}$ ($=120$ AU or $\sim 1"$), before dropping below the observed flux. This suggests that there is some material in the PDS 70 disk that is outside $R_{\rm disk}$ but is sufficiently tenuous that it does not contribute to the C$^{18}$O flux. In addition, the slope and peak of the $^{12}$CO may suggest that the temperature profile of the emitting layer is not consistent between model and observations. This could be caused by ignoring heating contribution from the embedded planets, or because our dust distribution is not providing the proper amount of opacity. The shifted peak suggests that the tenuous gas in the gap is warmer than we compute in our model. We leave the modelling of these aspects of the data to future work. For the purpose of understanding the planet formation within the disk the tenuous gas in the outer disk likely does not play an important role.

\subsubsection{C$_2$H}

\begin{figure*}
    \centering
    \includegraphics[width=\textwidth]{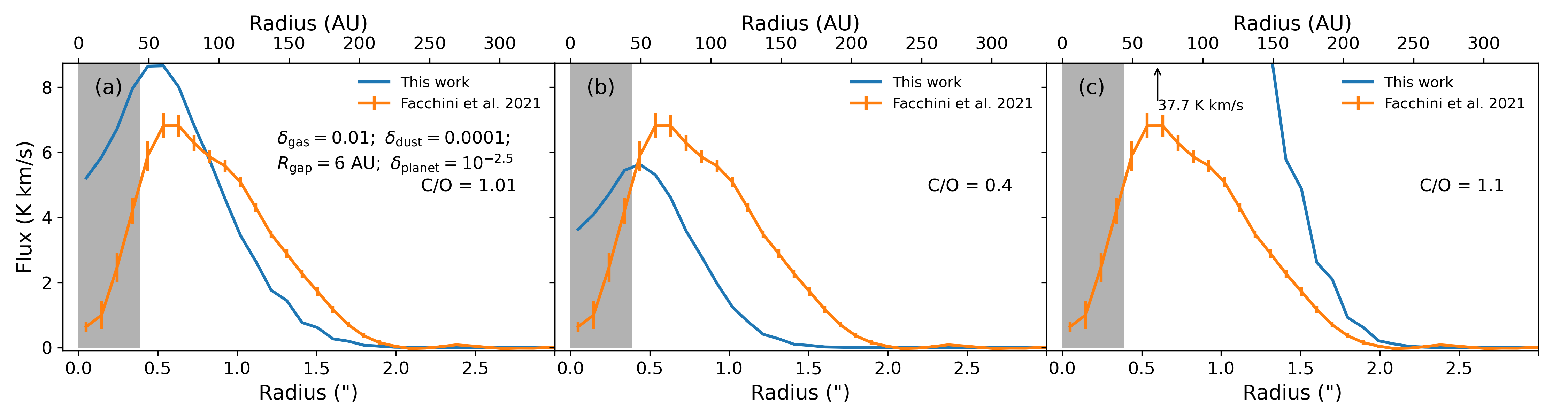}
    \caption{Radial profiles of C$_2$H emission compared to the work of \cite{Facchini2021}. In (a) we show the emission from the preferred disk model of Figure \ref{fig:triple_plot}(c), which has C/O $= 1.01$. The peak flux is slight over estimated, but it best reproduces the peak location compared to models with lower C/O. In (b) we show the same disk model, but with O/H $=2.525\times10^{-4}$ so that C/O = 0.4. Here we see that the peak flux is suppressed and the peak moves inward, in this model the inner disk dominates the C$_2$H flux. In (c) we set C/H $=1.1\times10^{-4}$ and O/H $=1.0\times10^{-4}$ such that C/O = 1.1. Here we find that the predicted flux greatly over estimates the observed C$_2$H flux. }
    \label{fig:triple_plot_C2H}
\end{figure*}

In Figure \ref{fig:triple_plot_C2H}(a) we show the resulting C$_2$H flux from the preferred model presented in Figure \ref{fig:triple_plot}(c). Our model predicts a similar peak flux - roughly 30\% larger than observed - with a similar peak location. Our model peak is shifted slightly inward, towards the inner disk, with respect to the observed peak location. This is due to bright C$_2$H emission that is found in the warm inner disk. This inner disk emission contributes to a bright C$_2$H line and regardless of our choice of disk parameters we find a similarly C$_2$H-enhanced inner disk.

Our modelling goal was to constrain the global carbon-to-oxygen in the PDS 70 disk. Our fiducial model was set such that the carbon abundance is that of the ISM, while the oxygen abundance is set lower so that C/O = 1.01. We expect a marginally carbon-rich disk based on the detection of C$_2$H, as well as a few other carbon-rich molecular species \citep{Facchini2021}. We test whether this hypothesis is robust by computing the C$_2$H emission for a disk with an ISM-like abundance of carbon and oxygen. 

In Figure \ref{fig:triple_plot_C2H}(b) we showed the preferred model from Figure \ref{fig:triple_plot}(c) but with an enhanced oxygen abundance such that the disk C/O $=0.4$. Not surprising, the total C$_2$H flux is suppressed in this model compared to the the marginally carbon-rich disk. Furthermore the flux peak shifts to smaller radii which suggests that the flux is being dominated by the inner disk in this model. For the low C/O model the carbon and oxygen abundances are set to their standard ISM abundances, while in the higher C/O models the oxygen abundances is depleted to O/H $=1.0\times 10^{-4}$, and the carbon abundance is adjusted to set the C/O ratios. Depleting abundances relative to the ISM value is consistent with previous studies of the emission of C$_2$H in protoplanetary disks \citep{Miotello2019,Bosman2021MAPS}, however their the abundances are depleted further than is done here. We further discuss our choice of carbon and oxygen abundances in the discussion section below.

In Figure \ref{fig:triple_plot_C2H}(c) we show the C$_2$H emission for a model with C/O $= 1.1$. The peak flux shifts further out and is more consistent with the location of the observed flux peak. However the model emission is about 4$\times$ brighter than is observed which heavily disfavours such carbon-rich models.

\begin{figure*}
    \centering
    \includegraphics[width=\textwidth]{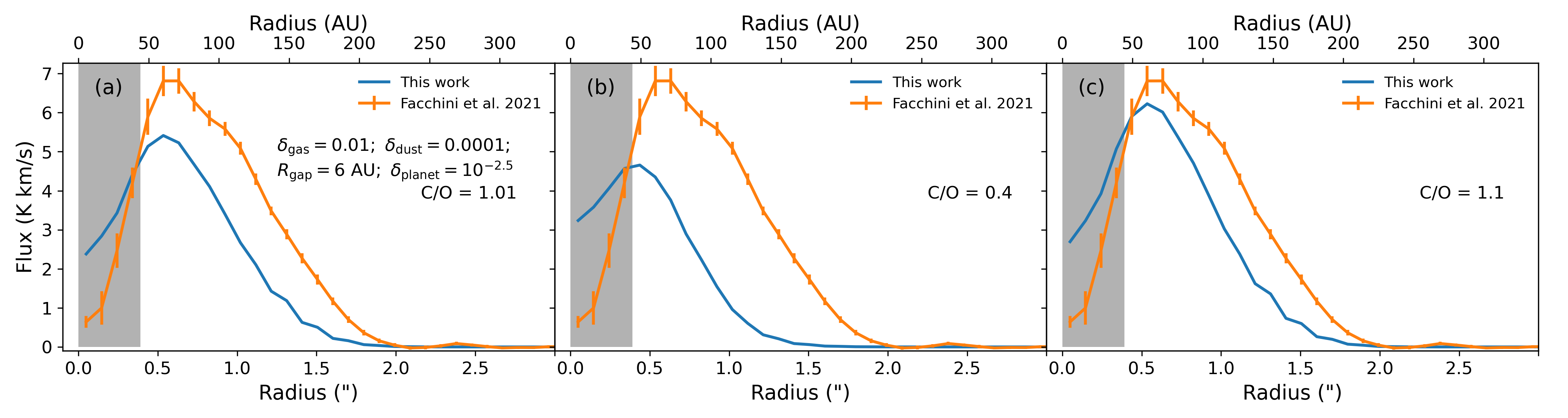}
    \caption{The same as figure \ref{fig:triple_plot_C2H} but for the model where C$_2$H is artificially removed in the inner 40 AU. The C/O $=0.4$ model (b) does not show a large change, however in both of the carbon-rich models we find that the inner disk is strongly contributing to the C$_2$H emission. Both the C/O $=1.01$ (a) and C/O $=1.1$ (c) models show similar C$_2$H flux profiles.}
    \label{fig:triple_plot_C2H_red}
\end{figure*}

In all three panels of figure \ref{fig:triple_plot_C2H} we find that the inner disk is bright in C$_2$H. In order to assess the impact of inner disk C$_2$H we artificially remove the C$_2$H from the inner 40 AU of each disk model and recalculate the radiative transfer of the C$_2$H emission, and shown the results in figure \ref{fig:triple_plot_C2H_red}. We find that in both of the carbon-rich models, panel (a) and (c), the removal of the inner disk C$_2$H emission has suppressed the total C$_2$H flux, while in the carbon-poor model (panel b) the profile is unaffected.

Our C$_2$H modelling has suggested that the inner disk and outer disk have different chemical properties. With the former showing evidence of having a lower elemental abundance than the outer disk due to its dominant C$_2$H flux in the preferred model. Future studies of PDS 70 may need to vary the elemental abundances as a function of radius, and higher spacial resolution studies of the gas emission will help to constrain this variation.

\subsubsection{Summary of chemical modelling}

In this section we have tested forward DALI models of disks using the derived outer disk parameters of \cite{Keppler2018,Keppler2019} with a grid of models that varied the properties of the inner disk. These models have shown that the outer disk parameters do an adequate job in reproducing the C$^{18}$O peak flux and peak location. In addition the peak $^{12}$CO emission is well reproduced, but is shifted with respect to the observed peak. This shift is consistent with previous findings of transitions disks \citep[see for ex.][]{vanderMarel2016} where the gas and dust cavity edges are shifted with respect to each other (although in our model they are the same), and could be linked to a more gradual decline in gas density at $R_{\rm cav}$ than we model here.

Using C$_2$H emission as a tracer of chemistry, we find that both the outer and inner disk contribute to its flux. For marginally carbon-rich models the flux peak is shifted towards the outer disk which is more consistent with the observed radial profile. Furthermore we show that for C/O $= 1.1$ the C$_2$H flux is far too strong to be consistent with observed emission, suggesting that the disk carbon and oxygen abundances are nearly equal, but slightly favours carbon. In our preferred model, we find that the inner disk is too bright in C$_2$H which may suggest a difference in the volatile C/O or C/H between the inner and outer disk. Such a chemical difference could be linked to the sequestering of either element into larger, trapped, dust grains as argued in \cite{Sturm2022}. Because of the high temperatures in the inner disk, the exact composition of the gas there can be directly probed by \textit{James Webb Space Telescope} (JWST).

In comparison to other studies of high-carbon protoplanetary disk chemistry we compute the integrated flux of C$_2$H and $^{13}$CO (whose azimuthially averaged distribution is shown in the appendix). We find that $\log_{10}(L_{\rm C_2H}) = 7.63$, in units of mJy km s$^{-1}$ pc$^2$, and $\log_{10}(F_{\rm ^{13}CO}/F_{\rm 890\mu m}) = 1.15$ which places our preferred model in a similar range of disks studied by \cite{Miotello2019}. For the total gas mass of the preferred model (at an age of 5 Myr), $1.5\times 10^{-3}$ M$_{\odot}$, we find that our predicted C$_2$H flux is consistent with the C/O models of \cite{Miotello2019} that are above unity. The column density of C$_2$H are on average higher than $10^{14}$ cm$^{-2}$ across the whole disk, which are similarly consistent with the models of \cite{Bosman2021MAPS} that have C/O above unity. Given what we find in comparing the radial distribution of the line emission and the comparisons to other works studying the C$_2$H flux in protoplanetary disks, we favour a marginally carbon-rich model for the current properties of the PDS 70 disk.

\subsection{Planet formation in the PDS 70 disk}

\subsubsection{Disk model at the time of planet formation}

\begin{figure}
    \centering
    \includegraphics[width=0.5\textwidth]{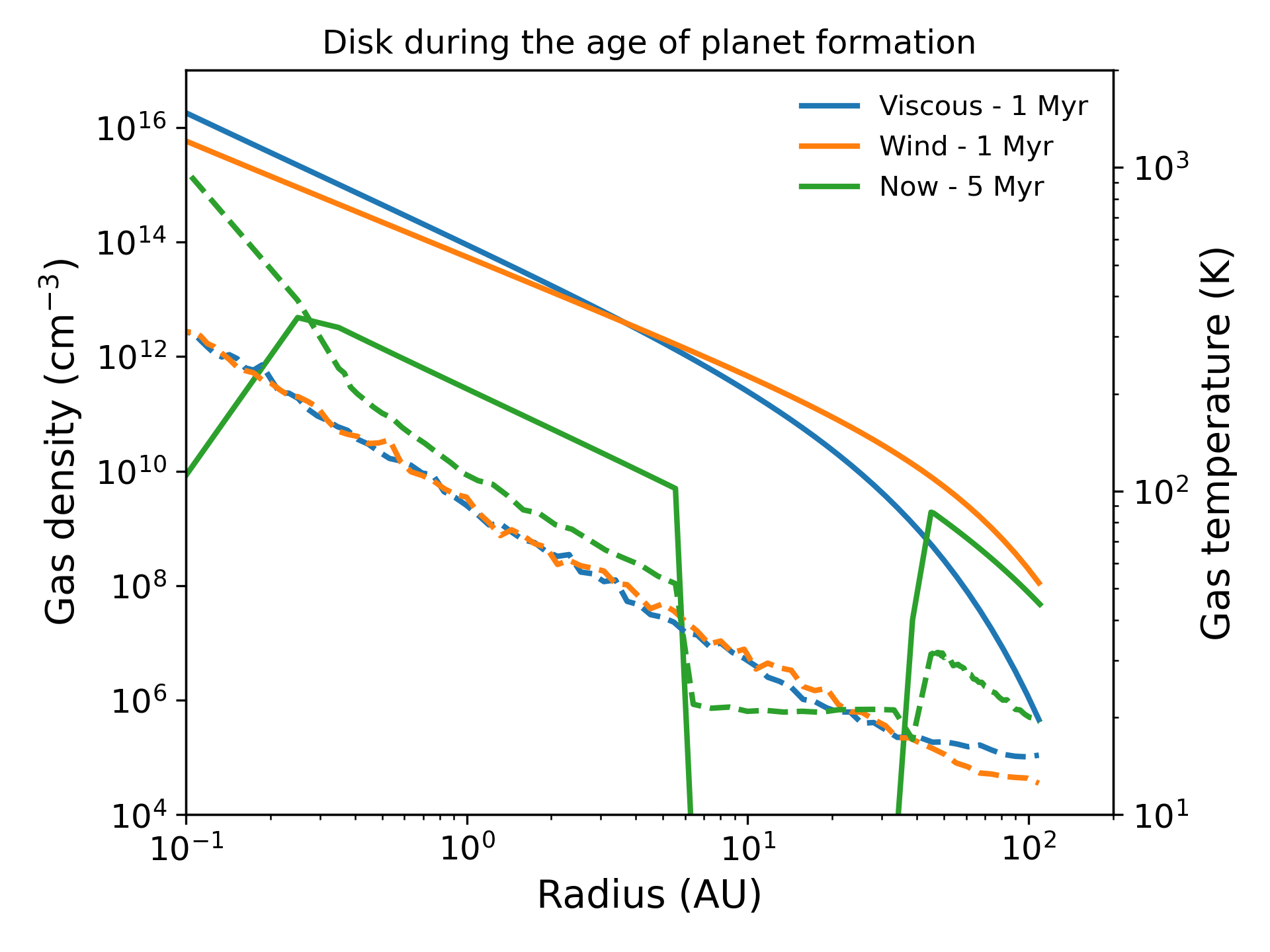}
    \caption{The midplane gas density profiles (solid lines) used for planet formation compared to the `current' (at 5 Myr) density profile (green). The `Wind' (orange) and `Viscous' (blue) models are related to the turnback models presented in Table \ref{tab:turnback} and Equation \ref{eq:tabone} at 1 Myr. The dashed lines show the gas temperature through each of the models. We assume that the disk temperature is dominated by the direct irradiation of the host star.}
    \label{fig:gas_profiles}
\end{figure}

In Figure \ref{fig:gas_profiles} we show the two gas density profiles (solid lines) determined by computing the evolution of the PDS 70 disk through viscous (blue) and disk wind (orange) evolution. Viscous disks must transport material outward to carry angular momentum away from the host star. As a result there is actually less gas in the outer region ($R>45$ AU) of the disk at a younger age than there is now (green). Wind-driven disks, on the other hand, transport angular momentum vertically and do not spread \citep{Tabone2021}. As a result its outer disk is more massive than both the outer gas disk today, as well as the viscous model.

The dashed lines in Figure \ref{fig:gas_profiles} show the midplane gas temperature through each of the models. The temperature profiles are similar between each of the disk models with the exception of the model representing the disk `now'. While disk models generally cool as they age, we are neglecting viscous heating in our chemical models, so the disk `now' is slightly warmer because the gas density has dropped, allowing for stellar irradiation deeper into the disk. By 1 Myr - our assumed age of the viscous and wind models - most evolving analytic models find that heating is dominated by direct stellar irradiation outside of a few AU \citep{Cham09}.

\subsubsection{Growth of the b and c planets}

\ignore{
\begin{figure}
    \centering
    \begin{minipage}{0.5\textwidth}
        \includegraphics[width=\textwidth]{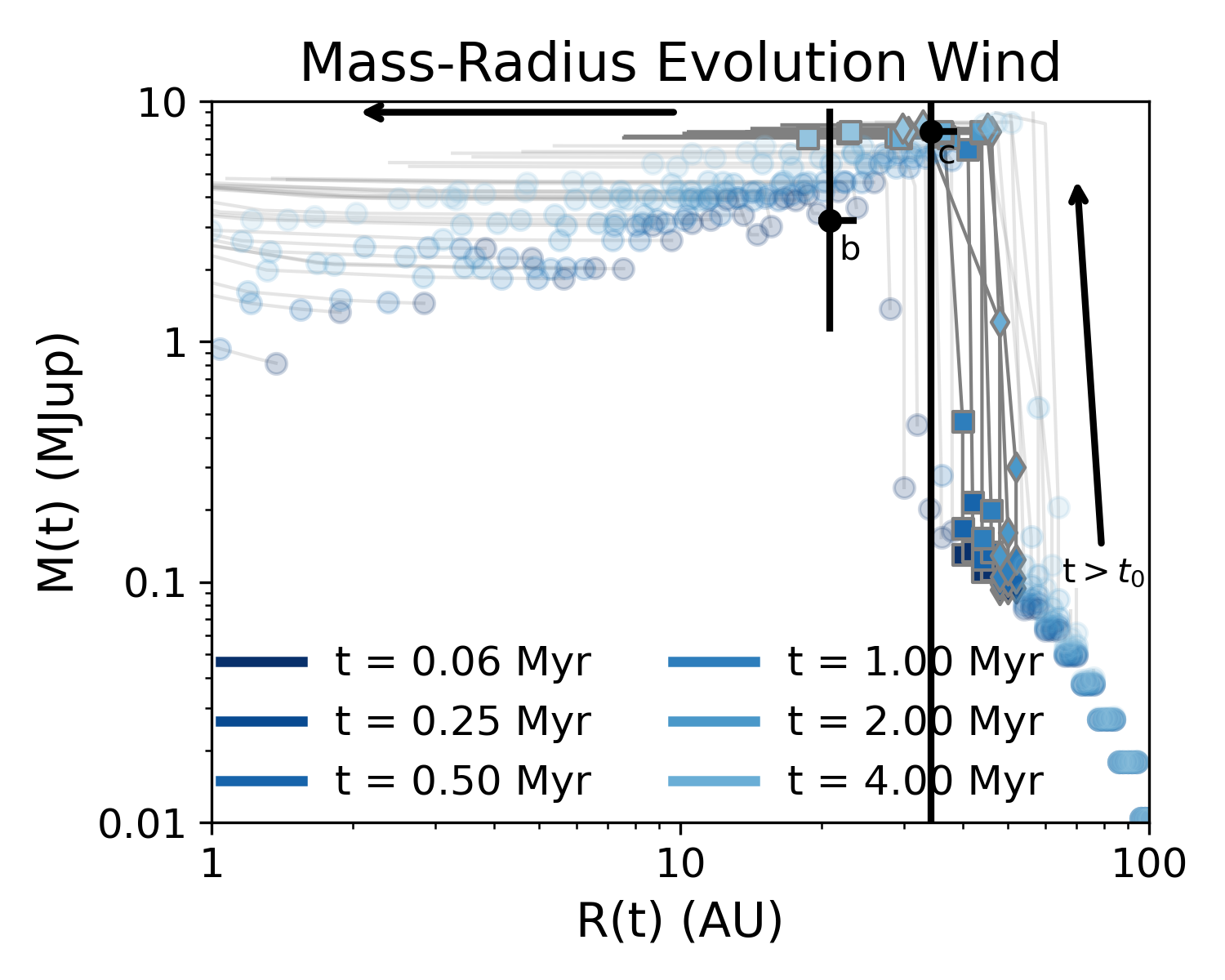}
        \caption{Evolution of synthetic planets in the disk model given by disk-wind evolution. The colour of each point denotes a particular timestep. The faded circles are synthetic planets that do not result in PDS 70-like planets at a time of 4 Myr after the beginning of the simulation. The square points show the successful PDS 70b synthetic planets, while the diamonds show the successful PDS 70c planets. The black circles and error bars are the values for PDS 70 b and c as shown in Table \ref{tab:the_planets}. The grey lines connect the points from individual synthetic planets. Each timestep shows the elapsed time since $t_0 = 1$ Myr.}
        \label{fig:mass-evo-wind}
    \end{minipage}
    \begin{minipage}{0.5\textwidth}
        \includegraphics[width=\textwidth]{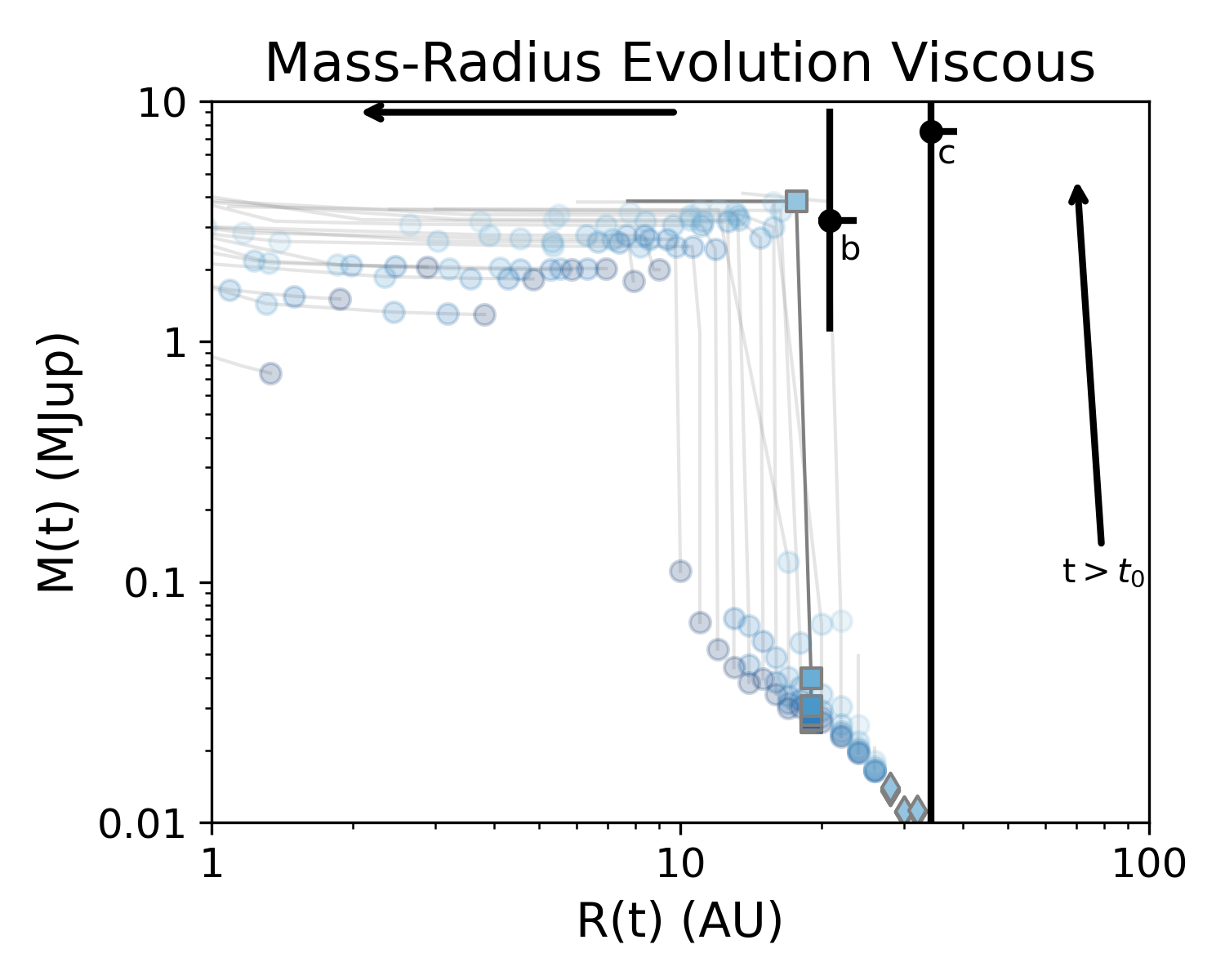}
        \caption{Same as in Figure \ref{fig:mass-evo-wind} but for the model that assumed the disk evolved through viscous torques. As stated in the text this disk density begins more centrally distributed and thus core accretion is less efficient in the outer disk in this model. Only PDS 70b can be represented by the population of synthetic planets, and only if twice the uncertainty on its position is used.}
        \label{fig:mass-evo-vis}
    \end{minipage}
\end{figure}
}

\begin{figure*}
    \centering
    \includegraphics[width=\textwidth]{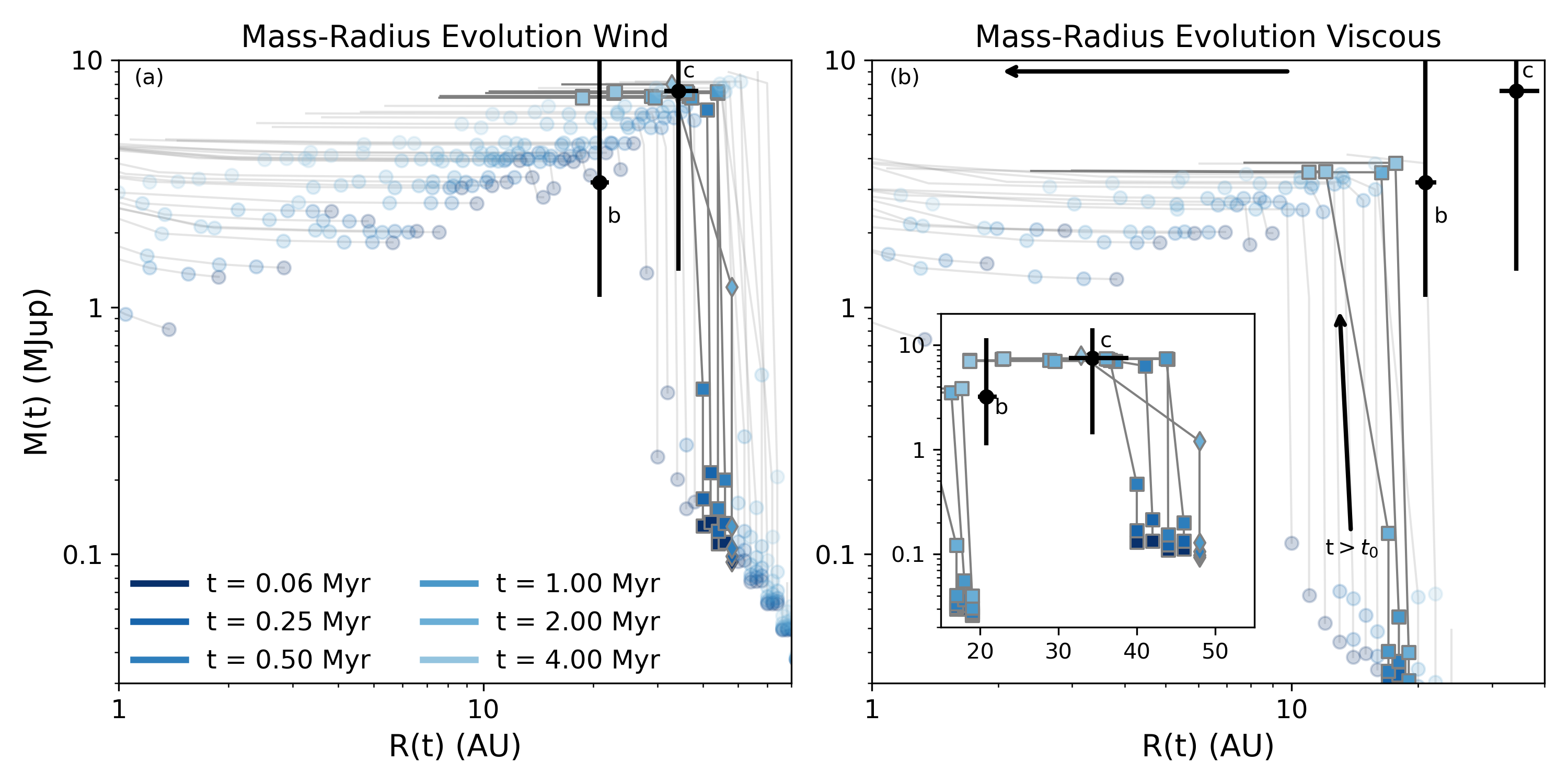}
    \caption{Evolution of synthetic planets in the disk model given by disk-wind evolution \textbf{(a)} and viscous evolution \textbf{(b)}. The colour of each point denotes a particular timestep since the begining of the simulation ($t_0=1$ Myr). The faded circles are synthetic planets that do not result in PDS 70-like planets at a time of 4 Myr after the beginning of the simulation. The square points show the successful PDS 70b synthetic planets, while the diamonds show the successful PDS 70c planets. The black circles and error bars are the values for PDS 70 b and c as shown in Table \ref{tab:the_planets}. The grey lines connect the points from individual synthetic planets. The wind model can reproduce both b and c planets, although it over estimates the mass of planet b. The viscous model can not reproduce the c planet and underestimates the radial location of the b planet. The inset of panel \textbf{(b)} shows the collection of successful synthetic planets for both the viscous and wind models. See the text for more details.}
    \label{fig:mass-evo}
\end{figure*}
Figures \ref{fig:mass-evo}\textbf{(a)} and \textbf{(b)} show the results of planet formation in the wind-driven and viscous-driven disk models respectively. In each figure the faded points denote synthetic planet formation tracks that fail to generate PDS 70-like planets, while the regular points are the successful simulations. The general evolution of synthetic planets is from the bottom and right of the figure towards the top left (shown with the black arrows - outward migration is not possible in our model) and thus the PDS70-like planets begin their evolution farther from PDS 70 than their current orbit.

The PDS 70-like synthetic planets have masses and orbital radii that are within twice the error of the measured values of PDS 70 b and c at a simulation age of 4 Myr. This simulation age corresponds to a system age of 5 Myr because we assume that planet formation begins at a system age of 1 Myr, and is consistent with the system properties inferred by \cite{Keppler2019}. Because of our choice of proximity within twice the uncertainty of the planets' currently known properties, in the viscous model (Figure \ref{fig:mass-evo}(b)) the PDS 70b-like planet actually starts near PDS 70b's current orbit and migrates slightly inward. This synthetic planet, however, is the only planet that has a similar mass and orbital radius as the PDS 70's current planets.

The drop in planet mass at all time steps in the outer regions of the disk are related to our choice of initial mass for the embryos. As the initial radius of the embryos moves outward there is less dust mass (because there is less disk exterior to its initial radius) and thus it starts at a lower mass than embryos beginning closer to the host star. The most extreme example of this is that planets forming inward of $\sim 30$ AU in the wind model, and $\sim 10$ AU in the viscous model have started sufficiently massive that their gas accretion already undergone a runaway by the first time step (60 kyr) shown. Meanwhile in the outer disk the embryos are not sufficiently heavy to undergo runaway gas accretion by 60 kyr, but do so later in the simulation.

Given the choice of system age we find that the Wind-driven model can better reproduce the mass and orbital radii of both planets than the viscous model. This is mainly linked to the fact that the viscous model requires that the density structure at 1 Myr is more centrally concentrated than the Wind model. As a result there is less material (both gas and dust) available in the outer disk for the construction of both the core and the gas envelope in the viscous model. Regardless, in both models we can identify a few of our synthetic planets that contain enough mass and orbits near to the PDS 70 planets' current orbit to investigate their resulting atmospheric chemistry.

In the wind-driven case we overestimate the mass of PDS 70b by a factor of about 2. We note here that, as in previous works, the planet formation of each synthetic planet is handled separately and thus the formation of the c-planet does not have any bearing on the formation of the b-planet. In reality, the growth of the c-planet likely impacted the gas flow from the disk to the b-planet, unless it formed much earlier than the c-planet. Our model has no sensitivity to differences in the formation start time, however this hints at other interesting lines of research for future studies of this system.

\subsection{Chemistry in the atmospheres of PDS 70b and c}

We compute the number of carbon and oxygen-carrying molecules that are available to both the b and c planets. We investigate two possible scenarios which are involved in setting the global C/O in the disk. The first is that the processes that lead to a high C/O (discussed above) happened very early in the lifetime of the PDS 70 system, and thus the global C/O is the same during the era of planet formation ($\sim 1$ Myr). On the other hand the processes that deplete oxygen and some carbon in the disk could have happened later in the disk lifetime, possibly after the planets had accreted most of their gas. In this scenario the global C/O would be the same as the stellar value $\sim 0.4$.

\begin{table}[]
    \centering
    \caption{The atmospheric carbon-to-oxygen ratios for the synthetic planets. }
    \begin{tabular}{c|c|c}
        \hline\hline
        Model & PDS 70b & PDS 70c\\\hline
        C/O = 1.01; Visc & 0.84 & \\
        C/O = 1.01; Wind & 0.99 (0.990-0.997) & 1.00 (0.999-1.001) \\
        C/O = 0.4; Visc & 0.43 & \\
        C/O = 0.4; Wind & 0.40 (0.402-0.403) & 0.40 (0.402-0.402) \\
        \hline
    \end{tabular}
    \label{tab:finalCtoO}
    \tablefoot{Each row denotes a different combination of disk model and chemical model. As seen in Figure \ref{fig:mass-evo} our viscous disk model fails to produce a satisfactory planet c. Where possible, the range of C/O from individual synthetic planets are shown to the 3rd decimal place.}
\end{table}

In Table \ref{tab:finalCtoO} we show the average C/O in all synthetic planets whose mass and orbital radii correspond to that of the PDS 70 pair of planets (within twice the uncertainties). Where multiple synthetic planets contributed to the average (i.e. in the Wind model) we show the range of C/O from each individual planet to an extra decimal place. The spread is small among separate planets. Not surprising, the planets resulting from the higher C/O models have higher C/O in their atmospheres. In the wind-driven model, where a satisfactory planet c can be found, the PDS 70b planet has a slightly lower C/O than the c-planet. Meanwhile the planet b results in a slightly different atmospheric C/O between the wind and viscous models. In both cases small differences in the formation history and the underlying chemical model lead to their varied atmospheric C/O. 

\begin{figure}
    \centering
    \includegraphics[width=0.5\textwidth]{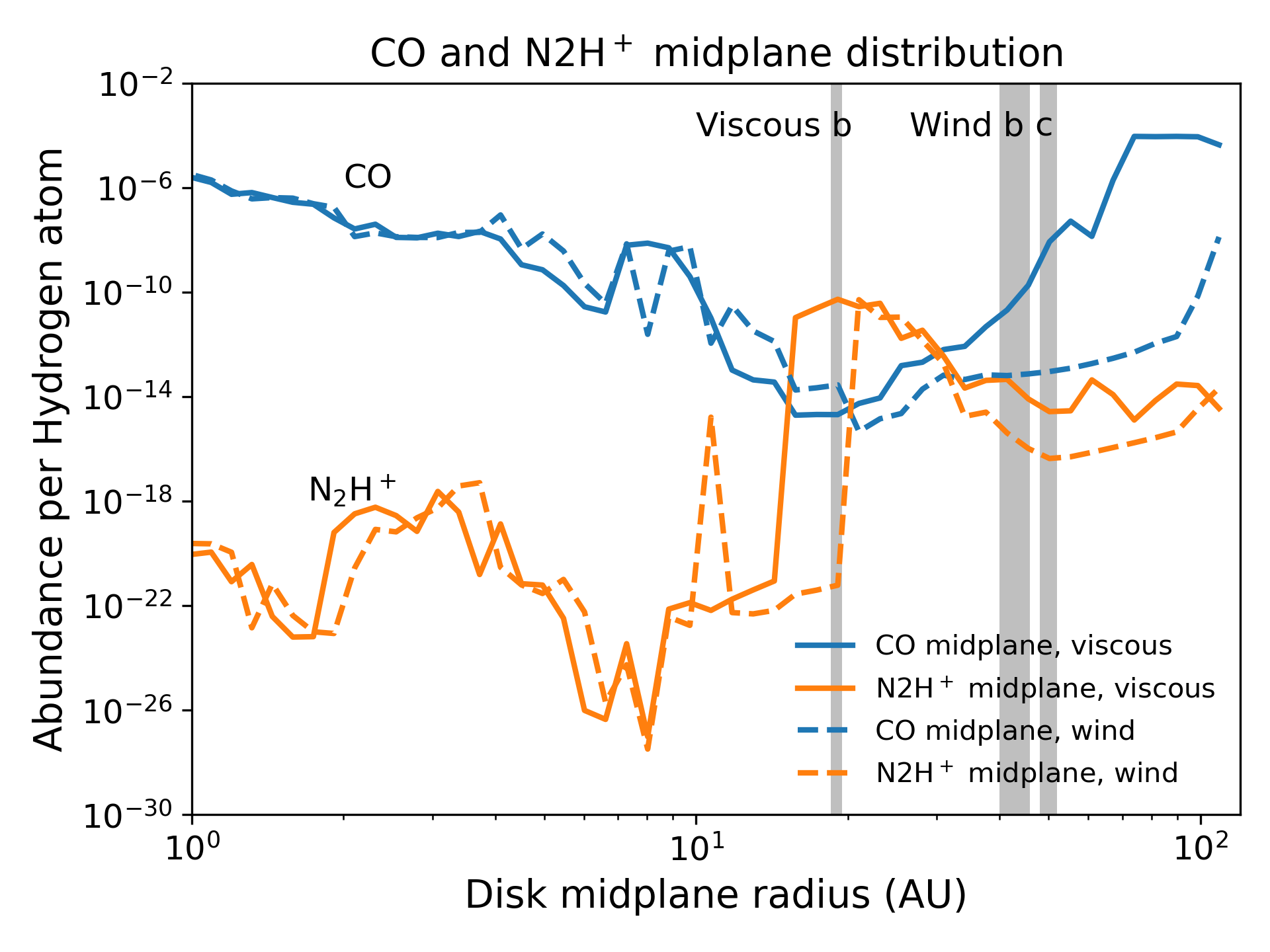}
    \caption{The midplane distribution of molecular abundances for CO and its tracer, N$_2$H$^+$ for both the viscous and wind models at 1 Myr. They each see a dip in CO abundance at the same position as a steep increase in N$_2$H$^+$, indicative of the CO ice line. The three grey bands show the range of initial embryo radii that result in the synthetic b and c planets in both the Viscous and Wind models.}
    \label{fig:COiceline}
\end{figure}

In Figure \ref{fig:COiceline} we show a representation of the CO ice line in our chemical models along the disk midplane. While the two-dimensional structure of the chemistry is important \citep[see for example ][]{Cridland2020}, for the purpose of the following discussion the midplane abundances can be illustrative.

The traditional definition of an ice line is the point in the disk where the temperature is such that the gas and ice phases of a species has equal abundances. Observationally, this is difficult to constrain and hence molecular tracers are used to infer the location of ice lines. For CO, N$_2$H$^+$ is a popular molecular tracer because its main destruction pathway involves reactions with CO \citep{Qi2013a,vtHoff2017}. In TW Hya, N$_2$H$^+$ has been used to infer the location of the CO ice line \citep{Qi2013b}.

The CO ice line is located between 15-30 AU along the midplane. Additionally on the figure the grey bars show the initial location of the embryos that led to the best PDS 70-like planets. They are, from left to right, PDS 70b in the viscous model, PDS 70b and PDS 70c in the wind model. By coincidence, we find that the b-planet in the viscous model begins its formation in the range of radii where the abundance of CO is near its minimum. Meanwhile the planets forming in the wind model form in a region where the CO abundances is higher. This higher abundance follows from higher interstellar radiation in the outer disk, which results in higher dust temperatures and thus less efficient freeze-out.

As argued in \cite{Cridland2020warmJs} the gas tends to be more carbon rich than the ices \citep[also see][]{Oberg11} and thus the relative amount of gas and solid abundance impacts the atmospheric C/O. In the outer disk, when CO begins to freeze out either as pure CO ice, or more likely as other \textit{oxygen-rich} ice species \citep[see for example ][]{Eistrup2018}, the bulk of the carbon and oxygen can become trapped in the ices. This is particularly important for the PDS 70b planet forming in the viscous model and can help explain its lower C/O compared to the wind model planets.

For the two planets forming in wind-driven model we find that their final masses are similar, and thus they acquire very similar atmospheric C/O. The (very) small difference in atmospheric C/O (at the level of $1\%$) may be due to the slight difference in their starting location and the resulting differences in the available CO in the gas phase. More available gaseous CO will lead to slightly higher atmospheric C/O. 

In the low C/O disk model we find that both PDS 70b and c show nearly equal atmospheric C/O in both disk models. Unlike in the high C/O models, planet formation in the outer disk appear to be less constrained by chemical gradients in the disk as the planet formation history of the planets.

\section{Discussion}\label{sec:discussion}

\subsection{Why study young systems?}\label{sec:discuss_youth}

There are several billion years between the point that a typical exoplanet's protoplanetary disk evaporates and the measurement of its atmospheric chemistry. Several dynamical processes, such as gravitational scattering between other planetary bodies (i.e. the Nice model, \cite{Nice}; or for hot Jupiters, \cite{Beauge2012}) and/or scattering by passing stars \citep{Shara2016,Hamers2017,Wang2020,Wang2022} could change the planet's orbital radius. These processes, however, should have no impact on the atmospheric chemistry if the planet has already fully formed and thus the aforementioned framework can still be used to interpret observations. 

Over billion year timescales, the mass-loss driven by photoevaporation can potentially change the chemistry of the upper atmosphere \citep{Yelle2004,GM2007,MC2009,Owen2012}. In the PDS 70 system, however these processes have limited effect on the chemistry of the two embedded planets. Firstly they orbit very far from their host stars compared to planets found in the well known hot Neptune desert \citep{Owen2018}, at orbital periods less than a few days, which are the main focus of photoevaporation studies. Furthermore, while heavy elements have been observed to be in the evaporating winds of exoplanets \citep{Fossati2010,Sing2019}, they usually remain coupled to the outflow \citep{Koskinen2013} which would maintain their \textit{relative} abundances in the remaining atmosphere \citep{Hunten1987}. Finally, the effect of photoevaporation on the chemical structure of giant planets likely does not play an important role on planets more massive than Saturn \citep{Mordasini16,Fossati2018}.

A final source of long timescale evolution that has the potential to change the observable chemical abundances of the atmospheres away from their primordial values are chemical reactions between the atmosphere and the planet's core. One possible direction for chemical evolution is through the envelope-induced core erosion that could transfer heavy elements from the core into the atmosphere via convection \citep{Stevenson1982,Stevenson1985,Guillot2004,Soubiran2017}. The study of this topic represents a rapidly developing field, however many studies have argued that the cores of gas giant planets consist of a diffuse outer and inner core with steep chemical gradients that suppress adiabatic convection in favour of less efficient heat and compositional transport from the core \citep{Stevenson1985,Chabrier2007,Leconte2012,Vazan2015,Vazan2016,Wahl2017,Moll2017,Vazan2018}.

In the other direction, differences in the average condensation temperatures of refractory (silicon, magnesium) and volatiles (oxygen, carbon) bearing species causes these species to rain-out at different altitudes. This can thus cause the chemical gradients in the planetary atmospheres - and its observable chemical abundances - to change as the planet loses its accretion energy \citep{Stevenson2022}. These internal processes can lead to difficulties in interpreting the chemical structure of old planets in the context of planet formation - depending on which molecular tracers are used. Young planets, on the other hand, still have the majority of their accretion heat and have not had enough time for many of these mixing processes to proceed. They thus provide an excellent test bed for studies linking the chemical properties of exoplanet atmospheres to planet formation physics. 

\subsection{Overview of the PDS 70 system}

The PDS 70 system offers a unique look at the planet formation process. While the star's metallicity is similar to that of the Sun, and so one would expect a stellar C/O that is near solar, the detection of many carbon-rich molecular features suggest that the current disk C/O is higher - perhaps above unity. The chemical modelling presented here seem to support this, however at low C/O the inner disk contributes strongly to the C$_2$H flux. This could be caused by two separate assumption regarding the chemistry in the disk which we discuss Section \ref{sec:ISMabun}.

Computing the classic core accretion scenario in the younger version of the PDS 70 disk has shown that super-Jupiter-massed planets at the orbital radii of the known planets is possible, however it is very sensitive to the accretion history of the disk. Viscously accreting disks - having had to begun more centrally compact - struggle to accumulate enough material in the outer disk to build both giant planets. Here we have assumed that the initial mass of the planetary embryos is the larger of the pebble isolation mass given by \cite{Bitsch2018} and the total quantity of solid mass exterior to the planet's starting location - assuming a gas-to-dust ratio of 100. As a result the initial core mass of our synthetic planets struggles to be large enough to draw down significant quantities of gas - particularly in the viscous model. 

We have considered two scenarios responsible for driving the disk's C/O away from the assumed stellar value. The first is that the responsible physical/chemical process occurs earlier in the disk life than the era of planet formation and thus the planets form in the same chemical environment that we see today (ie. high C/O). The second scenario imagines that the process leading to high C/O occurs late in the disk life, at least after the bulk of planet formation has already occurred and thus the planets form in a low C/O environment. Not surprisingly the resulting atmospheric C/O ratios are sensitive to this choice, with synthetic planets growing in the low C/O environment having themselves a low atmospheric C/O (effectively stellar) while in the opposite case the resulting C/O is super-stellar. 

Currently the measured C/O of PDS 70b and c are not well constrained, however the models of \cite{Wang2021} predict an atmospheric C/O for PDS 70b of between 0.54-0.7 depending on their choice of model. The sensitivity on the PDS 70c planet is worse than for PDS 70b, but using the same models \cite{Wang2021} find a range of C/O = 0.49-0.65. Clearly the current observations of the PDS 70 planets are not sufficiently sensitive to confidently differentiate between different models and future, more sensitive observations are required.

\subsection{ISM abundances of carbon}\label{sec:ISMabun}

In our chemical model we assume an ISM abundance of carbon, and modify the abundance of oxygen to set our disk C/O. We note that in previous work exploring the link between C$_2$H flux and volatile C/O, particularly \cite{Miotello2019} and \cite{Bosman2021MAPS}, have found that they must deplete the carbon and oxygen abundance by a factor of 100 relative to the ISM in order to better match observed and modelled column densities/fluxes. Here we find that in the case that we deplete the carbon and oxygen abundance we completely lose our C$^{18}$O flux that is coming from the inner edge of the outer dust ring (see Appendix \ref{sec:deplete}). We can therefore say that a global depletion of carbon and oxygen is not consistent with the observations of \cite{Facchini2021}.

In figure \ref{fig:most_abundant_spec} we show the most abundant carbon carriers for the model in figure \ref{fig:triple_plot}(a) and a model with a similar setup, but with carbon and oxygen abundances reduced by a factor of 100 \citep[ie. consistent with][]{Miotello2019} and a gas mass enhanced by a factor of 100. There we see that when the carbon and oxygen abundance is reduced the primary carbon carrier becomes CH$_4$ which freezes out onto the dust grains. This can be understood from a chemical kinetic perspective as follows: the production of CO scales with the abundance of both carbon and oxygen - and thus its production scales with $\rm n_C\dot n_O ~ n_X^2$, where $\rm X$ denotes the gas `metallicity'. Thus when the abundance of carbon and oxygen are reduced each by a factor of 100, the production of CO is slowed by a factor of 10000. CH$_4$, on the other hand is produced at a rate that scales with $\rm n_X$ - since it contains only one `heavy' element. It thus can become the dominant carbon carrier along the midplane of the disk at 5 Myr.

With that said, it is possible that \textit{local} depletions (or enhancements) of the carbon and oxygen are occurring in the PDS 70 disk. For example, there is a very obvious dust trap outwards of the orbital position of PDS 70c which could be locally enhanced in carbon and oxygen ices because of the ongoing flux of volatile rich ices there. Meanwhile the rest of the outer disk could be subsequently depleted in heavy elements. Furthermore the lack of many optically thin lines in the inner disk could suggest that it too is depleted in carbon and oxygen - particularly given that our chemical models consistently predict a high C$_2$H flux there, even at lower C/O. Further radially-dependent studies of the chemistry in the PDS 70 disk will be needed, along with upcoming high resolution ALMA data (Facchini et al. in prep) and new data from JWST.

\subsection{Choice of chemical network}

We followed the work of \cite{Miotello2019}, using their chemical network that was first developed by \cite{Visser2018}. The network contains 64 construction and destruction reactions for C$_2$H with a variety of reaction partners. The carbon chains are a maximum of two carbon atoms (ie. C$_2$H, C$_2$H$_2$, C$_2$H$_3$) which can result in erroneously large abundances of these species. 

As an illustration, \cite{Wei2019} computed the chemistry in protoplanetary disk for C/O greater than unity. They generally found that the carbon that was not incorporated into CO tended to inhabit long chain hydrocarbons and/or cyanides like HCN in their inner disk ($r <$ 5 AU). If these longer chain hydrocarbons were available in our chemical model it is possible that the inner disk chemistry may have shifted away from C$_2$H, but would have largely not affected the emission in the outer disk. We leave this investigation to future work. 

\subsection{Core accretion vs. gravitational instability}

In this work we have use the standard picture of core accretion to build our synthetic planets. As discussed above this planet formation scheme struggled at times to build planets similar to PDS 70c - at least in the case of a disk that evolved via viscous evolution. We have neglected the other popular planet formation mechanism that often leads to the generation of massive planets - gravitational instability \citep[GI, ie.][]{Boss1997}.

Very few astrochemical studies of GI currently exist, however \cite{Ilee2017} explored the chemical evolution of a pair of gravitationally unstable clumps in hydrodynamic simulations. They found that the clumps that survived throughout their whole simulation evolved to have the same atmospheric C/O as was initialised in their disk model. As such, we might expect that if PDS 70b and c formed through GI then they would have the same C/O as the disk. 

If this is true then it would be difficult to differentiate between the different formation models for PDS 70c with C/O alone - because its C/O seems to replicate the disk C/O here. In which case we may need to include an additional tracer of planet formation such as the solid to volatile ratio \citep{Schneider2021}, or the nitrogen abundance \citep{Bosman2019,Bitsch2022}. For PDS 70b there may be enough information encoded in the atmospheric C/O that can be used to differentiate the two formation mechanisms - at least if the disk C/O is high at the time of its formation.

In the future, it would be a useful to compute a similar experiment as was done in \cite{Ilee2017} to reproduce the PDS 70 pair of planets. Such a numerical simulation would also help to understand how the two planets impact their mutual growth and chemical evolution. In this work we ignore any mutual interaction between the two planets in their growth which is no doubt over-simplifying the physical system.

\section{Conclusion}\label{sec:conclusion}

We have modelled the chemical and physical structure of the PDS 70 disk in order to understand the environment in which the PDS 70b and PDS 70c planets formed. We find that the physical model of \cite{Keppler2018,Keppler2019} does an adequate job reproducing the C$^{18}$O and $^{12}$CO flux, and evidence of a tenuous inner disk. The carbon-to-oxygen ratio in the volatiles of the PDS 70 disk is likely marginally above unity, based on the flux of C$_2$H emission. The inner disk ($r< 10$ AU), however, is too bright in C$_2$H which may suggest that it is depleted in carbon and oxygen relative to the ISM-like abundances that we assume in this work. The current outer disk (at 5 Myr) is consistent with high abundances of carbon and oxygen, and we showed that if we deplete these elemental species we can not reproduce the flux of C$^{18}$O that comes from the inner edge of the outer dust ring.

To understand how the disk would have looked at the stage where planet formation began (at 1 Myr), we used analytic prescriptions for the evolution of the surface density and critical radius under the assumption that the disk evolved through viscous or MHD disk winds. The disk models resulting from the different driving mechanisms result in slightly different surface density profiles at an age of 1 Myr. The main difference between the two disk models is the quantity of material in the outer disk - which effects the formation of the synthetic planets around where the PDS 70 planets currently orbit.

We use a simple prescription for the growth of the PDS 70 planets, initialising their mass with the higher of the pebble isolation mass and the remaining total solid density exterior to the embryo's initial orbital radius. We accrete the gas and account for the collection of carbon and oxygen from both the gas and the volatile rich ices. Because it is unclear when the chemical processing of the volatiles occurred during the evolution of the disk we test two scenarios. In the case that the disk is chemically similar to it is now during the early era of planet formation the planets turn out to have a high C/O ratio, while in the opposite case - when the initial disk composition resembled the host star - the planets tended to have stellar C/O. The lower C/O models may be a better match to our current understanding of the planetary C/O, however the planetary C/O is only weakly constrained by observations which makes comparisons difficult to make.

The PDS 70 system represents a fantastic environment to study the link between planet formation and the planet's natal disk. In particular, understanding the chemical properties of the disk offer a unique opportunity to understand how giant planets acquire the higher mass elements like carbon and oxygen. This study will benefit greatly from both improved observational programs of the disk both at sub-mm wavelengths with ALMA as well as infrared with JWST, along with better constraints on the atmospheric C/O within the planets. Fortunately all of these studies are well on the way, and so PDS 70 will continue to be an excellent system to study planet formation for years to come.

\begin{acknowledgements}

 Thanks to the anonymous referee for their comments which greatly improved the clarity of this work. Thanks to the ExoGRAVITY team for stimulating discussion and observations that triggered this study. Astrochemistry in Leiden is supported by the Netherlands Research School for Astronomy (NOVA). AC and MB acknowledges funding from the European Research Council (ERC) under the European Union’s Horizon 2020 research and innovation programme (PROTOPLANETS, grant agreement No. 101002188). EFvD acknowledges support from the A-ERC, grant agreement No. 101019751 MOLDISK. SF is funded by the European Union under the European Union’s Horizon Europe Research \& Innovation Programme 101076613 (UNVEIL). This project made use of the following software: Astropy \citep{astropy}, SciPy \citep{Scipy}, NumPy \citep{numpy}, and Matplotlib \citep{matplotlib}. 

\end{acknowledgements}

\bibliographystyle{aa} 
\bibliography{mybib.bib} 

\begin{thebibliography}{129}
\expandafter\ifx\csname natexlab\endcsname\relax\def\natexlab#1{#1}\fi

\bibitem[{{Alessi} \& {Pudritz}(2018)}]{AP18}
{Alessi}, M. \& {Pudritz}, R.~E. 2018, \mnras, 478, 2599

\bibitem[{{Anderson} {et~al.}(2017){Anderson}, {Bergin}, {Blake}, {Ciesla},
  {Visser}, \& {Lee}}]{Anderson2017}
{Anderson}, D.~E., {Bergin}, E.~A., {Blake}, G.~A., {et~al.} 2017, \apj, 845,
  13

\bibitem[{{Andrews} {et~al.}(2011){Andrews}, {Wilner}, {Espaillat}, {Hughes},
  {Dullemond}, {McClure}, {Qi}, \& {Brown}}]{Andrews2011}
{Andrews}, S.~M., {Wilner}, D.~J., {Espaillat}, C., {et~al.} 2011, \apj, 732,
  42

\bibitem[{{Ansdell} {et~al.}(2016){Ansdell}, {Williams}, {van der Marel},
  {Carpenter}, {Guidi}, {Hogerheijde}, {Mathews}, {Manara}, {Miotello},
  {Natta}, {Oliveira}, {Tazzari}, {Testi}, {van Dishoeck}, \& {van
  Terwisga}}]{Ansdell2016}
{Ansdell}, M., {Williams}, J.~P., {van der Marel}, N., {et~al.} 2016, \apj,
  828, 46

\bibitem[{{Astropy Collaboration} {et~al.}(2022){Astropy Collaboration},
  {Price-Whelan}, {Lim}, {Earl}, {Starkman}, {Bradley}, {Shupe}, {Patil},
  {Corrales}, {Brasseur}, {N{\"o}the}, {Donath}, {Tollerud}, {Morris},
  {Ginsburg}, {Vaher}, {Weaver}, {Tocknell}, {Jamieson}, {van Kerkwijk},
  {Robitaille}, {Merry}, {Bachetti}, {G{\"u}nther}, {Aldcroft},
  {Alvarado-Montes}, {Archibald}, {B{\'o}di}, {Bapat}, {Barentsen},
  {Baz{\'a}n}, {Biswas}, {Boquien}, {Burke}, {Cara}, {Cara}, {Conroy},
  {Conseil}, {Craig}, {Cross}, {Cruz}, {D'Eugenio}, {Dencheva}, {Devillepoix},
  {Dietrich}, {Eigenbrot}, {Erben}, {Ferreira}, {Foreman-Mackey}, {Fox},
  {Freij}, {Garg}, {Geda}, {Glattly}, {Gondhalekar}, {Gordon}, {Grant},
  {Greenfield}, {Groener}, {Guest}, {Gurovich}, {Handberg}, {Hart},
  {Hatfield-Dodds}, {Homeier}, {Hosseinzadeh}, {Jenness}, {Jones}, {Joseph},
  {Kalmbach}, {Karamehmetoglu}, {Ka{\l}uszy{\'n}ski}, {Kelley}, {Kern},
  {Kerzendorf}, {Koch}, {Kulumani}, {Lee}, {Ly}, {Ma}, {MacBride}, {Maljaars},
  {Muna}, {Murphy}, {Norman}, {O'Steen}, {Oman}, {Pacifici}, {Pascual},
  {Pascual-Granado}, {Patil}, {Perren}, {Pickering}, {Rastogi}, {Roulston},
  {Ryan}, {Rykoff}, {Sabater}, {Sakurikar}, {Salgado}, {Sanghi}, {Saunders},
  {Savchenko}, {Schwardt}, {Seifert-Eckert}, {Shih}, {Jain}, {Shukla}, {Sick},
  {Simpson}, {Singanamalla}, {Singer}, {Singhal}, {Sinha}, {Sip{\H{o}}cz},
  {Spitler}, {Stansby}, {Streicher}, {{\v{S}}umak}, {Swinbank}, {Taranu},
  {Tewary}, {Tremblay}, {Val-Borro}, {Van Kooten}, {Vasovi{\'c}}, {Verma}, {de
  Miranda Cardoso}, {Williams}, {Wilson}, {Winkel}, {Wood-Vasey}, {Xue},
  {Yoachim}, {Zhang}, {Zonca}, \& {Astropy Project Contributors}}]{astropy}
{Astropy Collaboration}, {Price-Whelan}, A.~M., {Lim}, P.~L., {et~al.} 2022,
  \apj, 935, 167

\bibitem[{{Bae} {et~al.}(2019){Bae}, {Zhu}, {Baruteau}, {Benisty}, {Dullemond},
  {Facchini}, {Isella}, {Keppler}, {P{\'e}rez}, \& {Teague}}]{Bae2019}
{Bae}, J., {Zhu}, Z., {Baruteau}, C., {et~al.} 2019, \apjl, 884, L41

\bibitem[{{Batygin}(2018)}]{Batygin2018}
{Batygin}, K. 2018, \aj, 155, 178

\bibitem[{{Beaug{\'e}} \& {Nesvorn{\'y}}(2012)}]{Beauge2012}
{Beaug{\'e}}, C. \& {Nesvorn{\'y}}, D. 2012, \apj, 751, 119

\bibitem[{{Benisty} {et~al.}(2021){Benisty}, {Bae}, {Facchini}, {Keppler},
  {Teague}, {Isella}, {Kurtovic}, {P{\'e}rez}, {Sierra}, {Andrews},
  {Carpenter}, {Czekala}, {Dominik}, {Henning}, {Menard}, {Pinilla}, \&
  {Zurlo}}]{Benisty2021}
{Benisty}, M., {Bae}, J., {Facchini}, S., {et~al.} 2021, \apjl, 916, L2

\bibitem[{{Bergin} {et~al.}(2015){Bergin}, {Blake}, {Ciesla}, {Hirschmann}, \&
  {Li}}]{Berg15}
{Bergin}, E.~A., {Blake}, G.~A., {Ciesla}, F., {Hirschmann}, M.~M., \& {Li}, J.
  2015, Proceedings of the National Academy of Science, 112, 8965

\bibitem[{{Bergin} {et~al.}(2016){Bergin}, {Du}, {Cleeves}, {Blake}, {Schwarz},
  {Visser}, \& {Zhang}}]{Bergin2016}
{Bergin}, E.~A., {Du}, F., {Cleeves}, L.~I., {et~al.} 2016, \apj, 831, 101

\bibitem[{{Binkert} \& {Birnstiel}(2023)}]{BinkertBirnstiel2023}
{Binkert}, F. \& {Birnstiel}, T. 2023, \mnras, 520, 2055

\bibitem[{{Bitsch} {et~al.}(2018){Bitsch}, {Morbidelli}, {Johansen}, {Lega},
  {Lambrechts}, \& {Crida}}]{Bitsch2018}
{Bitsch}, B., {Morbidelli}, A., {Johansen}, A., {et~al.} 2018, \aap, 612, A30

\bibitem[{{Bitsch} {et~al.}(2022){Bitsch}, {Schneider}, \&
  {Kreidberg}}]{Bitsch2022}
{Bitsch}, B., {Schneider}, A.~D., \& {Kreidberg}, L. 2022, \aap, 665, A138

\bibitem[{{Booth} {et~al.}(2017){Booth}, {Clarke}, {Madhusudhan}, \&
  {Ilee}}]{Booth2017}
{Booth}, R.~A., {Clarke}, C.~J., {Madhusudhan}, N., \& {Ilee}, J.~D. 2017,
  \mnras, 469, 3994

\bibitem[{{Bosman} {et~al.}(2021){Bosman}, {Alarc{\'o}n}, {Bergin}, {Zhang},
  {van't Hoff}, {{\"O}berg}, {Guzm{\'a}n}, {Walsh}, {Aikawa}, {Andrews},
  {Bergner}, {Booth}, {Cataldi}, {Cleeves}, {Czekala}, {Furuya}, {Huang},
  {Ilee}, {Law}, {Le Gal}, {Liu}, {Long}, {Loomis}, {M{\'e}nard}, {Nomura},
  {Qi}, {Schwarz}, {Teague}, {Tsukagoshi}, {Yamato}, \&
  {Wilner}}]{Bosman2021MAPS}
{Bosman}, A.~D., {Alarc{\'o}n}, F., {Bergin}, E.~A., {et~al.} 2021, \apjs, 257,
  7

\bibitem[{{Bosman} {et~al.}(2019){Bosman}, {Cridland}, \&
  {Miguel}}]{Bosman2019}
{Bosman}, A.~D., {Cridland}, A.~J., \& {Miguel}, Y. 2019, \aap, 632, L11

\bibitem[{{Bosman} {et~al.}(2018{\natexlab{a}}){Bosman}, {Tielens}, \& {van
  Dishoeck}}]{Bosman2017b}
{Bosman}, A.~D., {Tielens}, A. G.~G.~M., \& {van Dishoeck}, E.~F.
  2018{\natexlab{a}}, \aap, 611, A80

\bibitem[{{Bosman} {et~al.}(2018{\natexlab{b}}){Bosman}, {Walsh}, \& {van
  Dishoeck}}]{Bosman2018}
{Bosman}, A.~D., {Walsh}, C., \& {van Dishoeck}, E.~F. 2018{\natexlab{b}},
  \aap, 618, A182

\bibitem[{{Boss}(1997)}]{Boss1997}
{Boss}, A.~P. 1997, Science, 276, 1836

\bibitem[{{Bruderer}(2013)}]{Bruderer2013}
{Bruderer}, S. 2013, \aap, 559, A46

\bibitem[{{Bruderer} {et~al.}(2012){Bruderer}, {van Dishoeck}, {Doty}, \&
  {Herczeg}}]{Bruderer2012}
{Bruderer}, S., {van Dishoeck}, E.~F., {Doty}, S.~D., \& {Herczeg}, G.~J. 2012,
  \aap, 541, A91

\bibitem[{{Chabrier} \& {Baraffe}(2007)}]{Chabrier2007}
{Chabrier}, G. \& {Baraffe}, I. 2007, \apjl, 661, L81

\bibitem[{{Chambers}(2009)}]{Cham09}
{Chambers}, J.~E. 2009, \apj, 705, 1206

\bibitem[{{Charnay} {et~al.}(2018){Charnay}, {B{\'e}zard}, {Baudino},
  {Bonnefoy}, {Boccaletti}, \& {Galicher}}]{Charnay2018}
{Charnay}, B., {B{\'e}zard}, B., {Baudino}, J.~L., {et~al.} 2018, \apj, 854,
  172

\bibitem[{{Christensen} {et~al.}(2009){Christensen}, {Holzwarth}, \&
  {Reiners}}]{Christensen2009}
{Christensen}, U.~R., {Holzwarth}, V., \& {Reiners}, A. 2009, \nat, 457, 167

\bibitem[{{Ciesla} \& {Cuzzi}(2006)}]{CieslaCuzzi2006}
{Ciesla}, F.~J. \& {Cuzzi}, J.~N. 2006, \icarus, 181, 178

\bibitem[{{Crida} {et~al.}(2006){Crida}, {Morbidelli}, \& {Masset}}]{Crida2006}
{Crida}, A., {Morbidelli}, A., \& {Masset}, F. 2006, \icarus, 181, 587

\bibitem[{{Cridland}(2018)}]{Cridland2018}
{Cridland}, A.~J. 2018, \aap, 619, A165

\bibitem[{{Cridland} {et~al.}(2020{\natexlab{a}}){Cridland}, {Bosman}, \& {van
  Dishoeck}}]{Cridland2020}
{Cridland}, A.~J., {Bosman}, A.~D., \& {van Dishoeck}, E.~F.
  2020{\natexlab{a}}, \aap, 635, A68

\bibitem[{{Cridland} {et~al.}(2016){Cridland}, {Pudritz}, \&
  {Alessi}}]{Crid16a}
{Cridland}, A.~J., {Pudritz}, R.~E., \& {Alessi}, M. 2016, \mnras, 461, 3274

\bibitem[{{Cridland} {et~al.}(2020{\natexlab{b}}){Cridland}, {van Dishoeck},
  {Alessi}, \& {Pudritz}}]{Cridland2020warmJs}
{Cridland}, A.~J., {van Dishoeck}, E.~F., {Alessi}, M., \& {Pudritz}, R.~E.
  2020{\natexlab{b}}, \aap, 642, A229

\bibitem[{{Cugno} {et~al.}(2021){Cugno}, {Patapis}, {Stolker}, {Quanz},
  {Boehle}, {Hoeijmakers}, {Marleau}, {Molli{\`e}re}, {Nasedkin}, \&
  {Snellen}}]{Cugno2021}
{Cugno}, G., {Patapis}, P., {Stolker}, T., {et~al.} 2021, \aap, 653, A12

\bibitem[{{Currie} {et~al.}(2022){Currie}, {Lawson}, {Schneider}, {Lyra},
  {Wisniewski}, {Grady}, {Guyon}, {Tamura}, {Kotani}, {Kawahara}, {Brandt},
  {Uyama}, {Muto}, {Dong}, {Kudo}, {Hashimoto}, {Fukagawa}, {Wagner}, {Lozi},
  {Chilcote}, {Tobin}, {Groff}, {Ward-Duong}, {Januszewski}, {Norris},
  {Tuthill}, {van der Marel}, {Sitko}, {Deo}, {Vievard}, {Jovanovic},
  {Martinache}, \& {Skaf}}]{Currie2022}
{Currie}, T., {Lawson}, K., {Schneider}, G., {et~al.} 2022, Nature Astronomy,
  6, 751

\bibitem[{{D'Angelo} {et~al.}(2010){D'Angelo}, {Durisen}, \&
  {Lissauer}}]{Dangelo2010}
{D'Angelo}, G., {Durisen}, R.~H., \& {Lissauer}, J.~J. 2010, {Giant Planet
  Formation}, ed. S.~{Seager}, 319--346

\bibitem[{{Du} {et~al.}(2015){Du}, {Bergin}, \& {Hogerheijde}}]{Du2015}
{Du}, F., {Bergin}, E.~A., \& {Hogerheijde}, M.~R. 2015, \apjl, 807, L32

\bibitem[{{D{\"u}rmann} \& {Kley}(2015)}]{DK15}
{D{\"u}rmann}, C. \& {Kley}, W. 2015, \aap, 574, A52

\bibitem[{{D{\"u}rmann} \& {Kley}(2017)}]{DK17}
{D{\"u}rmann}, C. \& {Kley}, W. 2017, \aap, 598, A80

\bibitem[{{Eistrup} {et~al.}(2016){Eistrup}, {Walsh}, \& {van
  Dishoeck}}]{Eistrup2016}
{Eistrup}, C., {Walsh}, C., \& {van Dishoeck}, E.~F. 2016, \aap, 595, A83

\bibitem[{{Eistrup} {et~al.}(2018){Eistrup}, {Walsh}, \& {van
  Dishoeck}}]{Eistrup2018}
{Eistrup}, C., {Walsh}, C., \& {van Dishoeck}, E.~F. 2018, \aap, 613, A14

\bibitem[{{Emsenhuber} {et~al.}(2021){Emsenhuber}, {Mordasini}, {Burn},
  {Alibert}, {Benz}, \& {Asphaug}}]{NGPPS1}
{Emsenhuber}, A., {Mordasini}, C., {Burn}, R., {et~al.} 2021, \aap, 656, A69

\bibitem[{{Facchini} {et~al.}(2021){Facchini}, {Teague}, {Bae}, {Benisty},
  {Keppler}, \& {Isella}}]{Facchini2021}
{Facchini}, S., {Teague}, R., {Bae}, J., {et~al.} 2021, \aj, 162, 99

\bibitem[{{Favre} {et~al.}(2013){Favre}, {Cleeves}, {Bergin}, {Qi}, \&
  {Blake}}]{Favre2013}
{Favre}, C., {Cleeves}, L.~I., {Bergin}, E.~A., {Qi}, C., \& {Blake}, G.~A.
  2013, \apjl, 776, L38

\bibitem[{{Fossati} {et~al.}(2010){Fossati}, {Haswell}, {Froning}, {Hebb},
  {Holmes}, {Kolb}, {Helling}, {Carter}, {Wheatley}, {Collier Cameron},
  {Loeillet}, {Pollacco}, {Street}, {Stempels}, {Simpson}, {Udry}, {Joshi},
  {West}, {Skillen}, \& {Wilson}}]{Fossati2010}
{Fossati}, L., {Haswell}, C.~A., {Froning}, C.~S., {et~al.} 2010, \apjl, 714,
  L222

\bibitem[{{Fossati} {et~al.}(2018){Fossati}, {Koskinen}, {Lothringer},
  {France}, {Young}, \& {Sreejith}}]{Fossati2018}
{Fossati}, L., {Koskinen}, T., {Lothringer}, J.~D., {et~al.} 2018, \apjl, 868,
  L30

\bibitem[{{Gaia Collaboration}(2020)}]{GaiaDR32020}
{Gaia Collaboration}. 2020, VizieR Online Data Catalog, I/350

\bibitem[{{Garc{\'\i}a Mu{\~n}oz}(2007)}]{GM2007}
{Garc{\'\i}a Mu{\~n}oz}, A. 2007, \planss, 55, 1426

\bibitem[{{Gomes} {et~al.}(2005){Gomes}, {Levison}, {Tsiganis}, \&
  {Morbidelli}}]{Nice}
{Gomes}, R., {Levison}, H.~F., {Tsiganis}, K., \& {Morbidelli}, A. 2005, \nat,
  435, 466

\bibitem[{{Gregorio-Hetem} {et~al.}(1992){Gregorio-Hetem}, {Lepine}, {Quast},
  {Torres}, \& {de La Reza}}]{GH1992}
{Gregorio-Hetem}, J., {Lepine}, J.~R.~D., {Quast}, G.~R., {Torres}, C.~A.~O.,
  \& {de La Reza}, R. 1992, \aj, 103, 549

\bibitem[{{Guillot} {et~al.}(2004){Guillot}, {Stevenson}, {Hubbard}, \&
  {Saumon}}]{Guillot2004}
{Guillot}, T., {Stevenson}, D.~J., {Hubbard}, W.~B., \& {Saumon}, D. 2004, in
  Jupiter. The Planet, Satellites and Magnetosphere, ed. F.~{Bagenal}, T.~E.
  {Dowling}, \& W.~B. {McKinnon}, Vol.~1, 35--57

\bibitem[{{Haffert} {et~al.}(2019){Haffert}, {Bohn}, {de Boer}, {Snellen},
  {Brinchmann}, {Girard}, {Keller}, \& {Bacon}}]{Haffert2019}
{Haffert}, S.~Y., {Bohn}, A.~J., {de Boer}, J., {et~al.} 2019, Nature
  Astronomy, 3, 749

\bibitem[{{Hamers} \& {Tremaine}(2017)}]{Hamers2017}
{Hamers}, A.~S. \& {Tremaine}, S. 2017, \aj, 154, 272

\bibitem[{{Hashimoto} {et~al.}(2012){Hashimoto}, {Dong}, {Kudo}, {Honda},
  {McClure}, {Zhu}, {Muto}, {Wisniewski}, {Abe}, {Brandner}, {Brandt},
  {Carson}, {Egner}, {Feldt}, {Fukagawa}, {Goto}, {Grady}, {Guyon}, {Hayano},
  {Hayashi}, {Hayashi}, {Henning}, {Hodapp}, {Ishii}, {Iye}, {Janson},
  {Kandori}, {Knapp}, {Kusakabe}, {Kuzuhara}, {Kwon}, {Matsuo}, {Mayama},
  {McElwain}, {Miyama}, {Morino}, {Moro-Martin}, {Nishimura}, {Pyo}, {Serabyn},
  {Suenaga}, {Suto}, {Suzuki}, {Takahashi}, {Takami}, {Takato}, {Terada},
  {Thalmann}, {Tomono}, {Turner}, {Watanabe}, {Yamada}, {Takami}, {Usuda}, \&
  {Tamura}}]{Hashimoto2012}
{Hashimoto}, J., {Dong}, R., {Kudo}, T., {et~al.} 2012, \apjl, 758, L19

\bibitem[{{Hashimoto} {et~al.}(2015){Hashimoto}, {Tsukagoshi}, {Brown}, {Dong},
  {Muto}, {Zhu}, {Wisniewski}, {Ohashi}, {kudo}, {Kusakabe}, {Abe}, {Akiyama},
  {Brandner}, {Brandt}, {Carson}, {Currie}, {Egner}, {Feldt}, {Grady}, {Guyon},
  {Hayano}, {Hayashi}, {Hayashi}, {Henning}, {Hodapp}, {Ishii}, {Iye},
  {Janson}, {Kandori}, {Knapp}, {Kuzuhara}, {Kwon}, {Matsuo}, {McElwain},
  {Mayama}, {Mede}, {Miyama}, {Morino}, {Moro-Martin}, {Nishimura}, {Pyo},
  {Serabyn}, {Suenaga}, {Suto}, {Suzuki}, {Takahashi}, {Takami}, {Takato},
  {Terada}, {Thalmann}, {Tomono}, {Turner}, {Watanabe}, {Yamada}, {Takami},
  {Usuda}, \& {Tamura}}]{Hashimoto2015}
{Hashimoto}, J., {Tsukagoshi}, T., {Brown}, J.~M., {et~al.} 2015, \apj, 799, 43

\bibitem[{{Heays} {et~al.}(2017){Heays}, {Bosman}, \& {van
  Dishoeck}}]{Heays2017}
{Heays}, A.~N., {Bosman}, A.~D., \& {van Dishoeck}, E.~F. 2017, \aap, 602, A105

\bibitem[{{Hoeijmakers} {et~al.}(2018){Hoeijmakers}, {Schwarz}, {Snellen}, {de
  Kok}, {Bonnefoy}, {Chauvin}, {Lagrange}, \& {Girard}}]{Hoeijmakers2018}
{Hoeijmakers}, H.~J., {Schwarz}, H., {Snellen}, I.~A.~G., {et~al.} 2018, \aap,
  617, A144

\bibitem[{{Hunten} {et~al.}(1987){Hunten}, {Pepin}, \& {Walker}}]{Hunten1987}
{Hunten}, D.~M., {Pepin}, R.~O., \& {Walker}, J.~C.~G. 1987, \icarus, 69, 532

\bibitem[{{Hunter}(2007)}]{matplotlib}
{Hunter}, J.~D. 2007, Computing in Science and Engineering, 9, 90

\bibitem[{{Ilee} {et~al.}(2017){Ilee}, {Forgan}, {Evans}, {Hall}, {Booth},
  {Clarke}, {Rice}, {Boley}, {Caselli}, {Hartquist}, \& {Rawlings}}]{Ilee2017}
{Ilee}, J.~D., {Forgan}, D.~H., {Evans}, M.~G., {et~al.} 2017, \mnras, 472, 189

\bibitem[{{Isella} {et~al.}(2019){Isella}, {Benisty}, {Teague}, {Bae},
  {Keppler}, {Facchini}, \& {P{\'e}rez}}]{Isella2019}
{Isella}, A., {Benisty}, M., {Teague}, R., {et~al.} 2019, \apjl, 879, L25

\bibitem[{{Kama} {et~al.}(2016){Kama}, {Bruderer}, {van Dishoeck},
  {Hogerheijde}, {Folsom}, {Miotello}, {Fedele}, {Belloche}, {G{\"u}sten}, \&
  {Wyrowski}}]{Kama2016b}
{Kama}, M., {Bruderer}, S., {van Dishoeck}, E.~F., {et~al.} 2016, \aap, 592,
  A83

\bibitem[{{Keppler} {et~al.}(2018){Keppler}, {Benisty}, {M{\"u}ller},
  {Henning}, {van Boekel}, {Cantalloube}, {Ginski}, {van Holstein}, {Maire},
  {Pohl}, {Samland}, {Avenhaus}, {Baudino}, {Boccaletti}, {de Boer},
  {Bonnefoy}, {Chauvin}, {Desidera}, {Langlois}, {Lazzoni}, {Marleau},
  {Mordasini}, {Pawellek}, {Stolker}, {Vigan}, {Zurlo}, {Birnstiel},
  {Brandner}, {Feldt}, {Flock}, {Girard}, {Gratton}, {Hagelberg}, {Isella},
  {Janson}, {Juhasz}, {Kemmer}, {Kral}, {Lagrange}, {Launhardt}, {Matter},
  {M{\'e}nard}, {Milli}, {Molli{\`e}re}, {Olofsson}, {P{\'e}rez}, {Pinilla},
  {Pinte}, {Quanz}, {Schmidt}, {Udry}, {Wahhaj}, {Williams}, {Buenzli},
  {Cudel}, {Dominik}, {Galicher}, {Kasper}, {Lannier}, {Mesa}, {Mouillet},
  {Peretti}, {Perrot}, {Salter}, {Sissa}, {Wildi}, {Abe}, {Antichi},
  {Augereau}, {Baruffolo}, {Baudoz}, {Bazzon}, {Beuzit}, {Blanchard}, {Brems},
  {Buey}, {De Caprio}, {Carbillet}, {Carle}, {Cascone}, {Cheetham}, {Claudi},
  {Costille}, {Delboulb{\'e}}, {Dohlen}, {Fantinel}, {Feautrier}, {Fusco},
  {Giro}, {Gluck}, {Gry}, {Hubin}, {Hugot}, {Jaquet}, {Le Mignant}, {Llored},
  {Madec}, {Magnard}, {Martinez}, {Maurel}, {Meyer}, {M{\"o}ller-Nilsson},
  {Moulin}, {Mugnier}, {Orign{\'e}}, {Pavlov}, {Perret}, {Petit}, {Pragt},
  {Puget}, {Rabou}, {Ramos}, {Rigal}, {Rochat}, {Roelfsema}, {Rousset}, {Roux},
  {Salasnich}, {Sauvage}, {Sevin}, {Soenke}, {Stadler}, {Suarez}, {Turatto}, \&
  {Weber}}]{Keppler2018}
{Keppler}, M., {Benisty}, M., {M{\"u}ller}, A., {et~al.} 2018, \aap, 617, A44

\bibitem[{{Keppler} {et~al.}(2019){Keppler}, {Teague}, {Bae}, {Benisty},
  {Henning}, {van Boekel}, {Chapillon}, {Pinilla}, {Williams}, {Bertrang},
  {Facchini}, {Flock}, {Ginski}, {Juhasz}, {Klahr}, {Liu}, {M{\"u}ller},
  {P{\'e}rez}, {Pohl}, {Rosotti}, {Samland}, \& {Semenov}}]{Keppler2019}
{Keppler}, M., {Teague}, R., {Bae}, J., {et~al.} 2019, \aap, 625, A118

\bibitem[{{Klarmann} {et~al.}(2018){Klarmann}, {Ormel}, \&
  {Dominik}}]{Klarmann2018}
{Klarmann}, L., {Ormel}, C.~W., \& {Dominik}, C. 2018, \aap, 618, L1

\bibitem[{{Kordopatis} {et~al.}(2013){Kordopatis}, {Gilmore}, {Steinmetz},
  {Boeche}, {Seabroke}, {Siebert}, {Zwitter}, {Binney}, {de Laverny},
  {Recio-Blanco}, {Williams}, {Piffl}, {Enke}, {Roeser}, {Bijaoui}, {Wyse},
  {Freeman}, {Munari}, {Carrillo}, {Anguiano}, {Burton}, {Campbell}, {Cass},
  {Fiegert}, {Hartley}, {Parker}, {Reid}, {Ritter}, {Russell}, {Stupar},
  {Watson}, {Bienaym{\'e}}, {Bland-Hawthorn}, {Gerhard}, {Gibson}, {Grebel},
  {Helmi}, {Navarro}, {Conrad}, {Famaey}, {Faure}, {Just}, {Kos},
  {Matijevi{\v{c}}}, {McMillan}, {Minchev}, {Scholz}, {Sharma}, {Siviero}, {de
  Boer}, \& {{\v{Z}}erjal}}]{Rave4}
{Kordopatis}, G., {Gilmore}, G., {Steinmetz}, M., {et~al.} 2013, \aj, 146, 134

\bibitem[{{Koskinen} {et~al.}(2013){Koskinen}, {Harris}, {Yelle}, \&
  {Lavvas}}]{Koskinen2013}
{Koskinen}, T.~T., {Harris}, M.~J., {Yelle}, R.~V., \& {Lavvas}, P. 2013,
  \icarus, 226, 1678

\bibitem[{{Krijt} {et~al.}(2020){Krijt}, {Bosman}, {Zhang}, {Schwarz},
  {Ciesla}, \& {Bergin}}]{Krijt2020}
{Krijt}, S., {Bosman}, A.~D., {Zhang}, K., {et~al.} 2020, \apj, 899, 134

\bibitem[{{Leconte} \& {Chabrier}(2012)}]{Leconte2012}
{Leconte}, J. \& {Chabrier}, G. 2012, \aap, 540, A20

\bibitem[{{Leemker} {et~al.}(2022){Leemker}, {Booth}, {van Dishoeck},
  {P{\'e}rez-S{\'a}nchez}, {Szul{\'a}gyi}, {Bosman}, {Bruderer}, {Facchini},
  {Hogerheijde}, {Paneque-Carre{\~n}o}, \& {Sturm}}]{Leemker2022}
{Leemker}, M., {Booth}, A.~S., {van Dishoeck}, E.~F., {et~al.} 2022, \aap, 663,
  A23

\bibitem[{{Lin} \& {Papaloizou}(1986)}]{LP86}
{Lin}, D.~N.~C. \& {Papaloizou}, J. 1986, \apj, 309, 846

\bibitem[{{Long} {et~al.}(2018){Long}, {Akiyama}, {Sitko}, {Fernandes},
  {Assani}, {Grady}, {Cure}, {Danchi}, {Dong}, {Fukagawa}, {Hasegawa},
  {Hashimoto}, {Henning}, {Inutsuka}, {Kraus}, {Kwon}, {Lisse}, {Baobabu Liu},
  {Mayama}, {Muto}, {Nakagawa}, {Takami}, {Tamura}, {Currie}, {Wisniewski}, \&
  {Yang}}]{Long2018}
{Long}, Z.~C., {Akiyama}, E., {Sitko}, M., {et~al.} 2018, \apj, 858, 112

\bibitem[{{Lubow} \& {D'Angelo}(2006)}]{Lubow2006}
{Lubow}, S.~H. \& {D'Angelo}, G. 2006, \apj, 641, 526

\bibitem[{{Lynden-Bell} \& {Pringle}(1974)}]{LB74}
{Lynden-Bell}, D. \& {Pringle}, J.~E. 1974, \mnras, 168, 603

\bibitem[{{Madhusudhan}(2019)}]{Madhu2019}
{Madhusudhan}, N. 2019, \araa, 57, 617

\bibitem[{{McElroy} {et~al.}(2013){McElroy}, {Walsh}, {Markwick}, {Cordiner},
  {Smith}, \& {Millar}}]{McE03}
{McElroy}, D., {Walsh}, C., {Markwick}, A.~J., {et~al.} 2013, \aap, 550, A36

\bibitem[{{Metchev} {et~al.}(2004){Metchev}, {Hillenbrand}, \&
  {Meyer}}]{MSH2004}
{Metchev}, S.~A., {Hillenbrand}, L.~A., \& {Meyer}, M.~R. 2004, \apj, 600, 435

\bibitem[{{Miotello} {et~al.}(2019){Miotello}, {Facchini}, {van Dishoeck},
  {Cazzoletti}, {Testi}, {Williams}, {Ansdell}, {van Terwisga}, \& {van der
  Marel}}]{Miotello2019}
{Miotello}, A., {Facchini}, S., {van Dishoeck}, E.~F., {et~al.} 2019, \aap,
  631, A69

\bibitem[{{Miotello} {et~al.}(2022){Miotello}, {Kamp}, {Birnstiel}, {Cleeves},
  \& {Kataoka}}]{Miotello2022}
{Miotello}, A., {Kamp}, I., {Birnstiel}, T., {Cleeves}, L.~I., \& {Kataoka}, A.
  2022, arXiv e-prints, arXiv:2203.09818

\bibitem[{{Miotello} {et~al.}(2014){Miotello}, {Testi}, {Lodato}, {Ricci},
  {Rosotti}, {Brooks}, {Maury}, \& {Natta}}]{Miotello2014}
{Miotello}, A., {Testi}, L., {Lodato}, G., {et~al.} 2014, \aap, 567, A32

\bibitem[{{Miotello} {et~al.}(2016){Miotello}, {van Dishoeck}, {Kama}, \&
  {Bruderer}}]{Miotello2016}
{Miotello}, A., {van Dishoeck}, E.~F., {Kama}, M., \& {Bruderer}, S. 2016,
  Astronomy and Astrophysics, 594, A85

\bibitem[{{Moll} {et~al.}(2017){Moll}, {Garaud}, {Mankovich}, \&
  {Fortney}}]{Moll2017}
{Moll}, R., {Garaud}, P., {Mankovich}, C., \& {Fortney}, J.~J. 2017, \apj, 849,
  24

\bibitem[{{Morbidelli} {et~al.}(2014){Morbidelli}, {Szul{\'a}gyi}, {Crida},
  {Lega}, {Bitsch}, {Tanigawa}, \& {Kanagawa}}]{Morbidelli2014}
{Morbidelli}, A., {Szul{\'a}gyi}, J., {Crida}, A., {et~al.} 2014, \icarus, 232,
  266

\bibitem[{{Mordasini} {et~al.}(2014){Mordasini}, {Klahr}, {Alibert}, {Miller},
  \& {Henning}}]{Mordasini14}
{Mordasini}, C., {Klahr}, H., {Alibert}, Y., {Miller}, N., \& {Henning}, T.
  2014, \aap, 566, A141

\bibitem[{{Mordasini} {et~al.}(2015){Mordasini}, {Molli{\`e}re}, {Dittkrist},
  {Jin}, \& {Alibert}}]{Mordasini15}
{Mordasini}, C., {Molli{\`e}re}, P., {Dittkrist}, K.-M., {Jin}, S., \&
  {Alibert}, Y. 2015, International Journal of Astrobiology, 14, 201

\bibitem[{{Mordasini} {et~al.}(2016){Mordasini}, {van Boekel}, {Molli{\`e}re},
  {Henning}, \& {Benneke}}]{Mordasini16}
{Mordasini}, C., {van Boekel}, R., {Molli{\`e}re}, P., {Henning}, T., \&
  {Benneke}, B. 2016, \apj, 832, 41

\bibitem[{{Murray-Clay} {et~al.}(2009){Murray-Clay}, {Chiang}, \&
  {Murray}}]{MC2009}
{Murray-Clay}, R.~A., {Chiang}, E.~I., \& {Murray}, N. 2009, \apj, 693, 23

\bibitem[{{{\"O}berg} {et~al.}(2021){{\"O}berg}, {Guzm{\'a}n}, {Walsh},
  {Aikawa}, {Bergin}, {Law}, {Loomis}, {Alarc{\'o}n}, {Andrews}, {Bae},
  {Bergner}, {Boehler}, {Booth}, {Bosman}, {Calahan}, {Cataldi}, {Cleeves},
  {Czekala}, {Furuya}, {Huang}, {Ilee}, {Kurtovic}, {Le Gal}, {Liu}, {Long},
  {M{\'e}nard}, {Nomura}, {P{\'e}rez}, {Qi}, {Schwarz}, {Sierra}, {Teague},
  {Tsukagoshi}, {Yamato}, {van't Hoff}, {Waggoner}, {Wilner}, \&
  {Zhang}}]{Oberg2021}
{{\"O}berg}, K.~I., {Guzm{\'a}n}, V.~V., {Walsh}, C., {et~al.} 2021, \apjs,
  257, 1

\bibitem[{{{\"O}berg} {et~al.}(2011){{\"O}berg}, {Murray-Clay}, \&
  {Bergin}}]{Oberg11}
{{\"O}berg}, K.~I., {Murray-Clay}, R., \& {Bergin}, E.~A. 2011, \apjl, 743, L16

\bibitem[{{Owen} \& {Jackson}(2012)}]{Owen2012}
{Owen}, J.~E. \& {Jackson}, A.~P. 2012, \mnras, 425, 2931

\bibitem[{{Owen} \& {Lai}(2018)}]{Owen2018}
{Owen}, J.~E. \& {Lai}, D. 2018, \mnras, 479, 5012

\bibitem[{{Pollack}(1984)}]{Pollack1984}
{Pollack}, J.~B. 1984, \araa, 22, 389

\bibitem[{{Pollack} {et~al.}(1996){Pollack}, {Hubickyj}, {Bodenheimer},
  {Lissauer}, {Podolak}, \& {Greenzweig}}]{Pollack1996}
{Pollack}, J.~B., {Hubickyj}, O., {Bodenheimer}, P., {et~al.} 1996, \icarus,
  124, 62

\bibitem[{{Qi} {et~al.}(2013{\natexlab{a}}){Qi}, {{\"O}berg}, \&
  {Wilner}}]{Qi2013a}
{Qi}, C., {{\"O}berg}, K.~I., \& {Wilner}, D.~J. 2013{\natexlab{a}}, \apj, 765,
  34

\bibitem[{{Qi} {et~al.}(2013{\natexlab{b}}){Qi}, {{\"O}berg}, {Wilner},
  {D'Alessio}, {Bergin}, {Andrews}, {Blake}, {Hogerheijde}, \& {van
  Dishoeck}}]{Qi2013b}
{Qi}, C., {{\"O}berg}, K.~I., {Wilner}, D.~J., {et~al.} 2013{\natexlab{b}},
  Science, 341, 630

\bibitem[{{Riaud} {et~al.}(2006){Riaud}, {Mawet}, {Absil}, {Boccaletti},
  {Baudoz}, {Herwats}, \& {Surdej}}]{Riaud2006}
{Riaud}, P., {Mawet}, D., {Absil}, O., {et~al.} 2006, \aap, 458, 317

\bibitem[{{Robert} {et~al.}(2018){Robert}, {Crida}, {Lega}, {M{\'e}heut}, \&
  {Morbidelli}}]{Robert2018}
{Robert}, C.~M.~T., {Crida}, A., {Lega}, E., {M{\'e}heut}, H., \& {Morbidelli},
  A. 2018, \aap, 617, A98

\bibitem[{{Schneider} \& {Bitsch}(2021)}]{Schneider2021}
{Schneider}, A.~D. \& {Bitsch}, B. 2021, \aap, 654, A72

\bibitem[{{Schwarz} {et~al.}(2018){Schwarz}, {Bergin}, {Cleeves}, {Zhang},
  {{\"O}berg}, {Blake}, \& {Anderson}}]{Schwarz2018}
{Schwarz}, K.~R., {Bergin}, E.~A., {Cleeves}, L.~I., {et~al.} 2018, \apj, 856,
  85

\bibitem[{{Shakura} \& {Sunyaev}(1973)}]{SS73}
{Shakura}, N.~I. \& {Sunyaev}, R.~A. 1973, \aap, 24, 337

\bibitem[{{Shara} {et~al.}(2016){Shara}, {Hurley}, \& {Mardling}}]{Shara2016}
{Shara}, M.~M., {Hurley}, J.~R., \& {Mardling}, R.~A. 2016, \apj, 816, 59

\bibitem[{{Sing} {et~al.}(2019){Sing}, {Lavvas}, {Ballester}, {Lecavelier des
  Etangs}, {Marley}, {Nikolov}, {Ben-Jaffel}, {Bourrier}, {Buchhave}, {Deming},
  {Ehrenreich}, {Mikal-Evans}, {Kataria}, {Lewis}, {L{\'o}pez-Morales},
  {Garc{\'\i}a Mu{\~n}oz}, {Henry}, {Sanz-Forcada}, {Spake}, {Wakeford}, \&
  {PanCET Collaboration}}]{Sing2019}
{Sing}, D.~K., {Lavvas}, P., {Ballester}, G.~E., {et~al.} 2019, \aj, 158, 91

\bibitem[{{Soubiran} {et~al.}(2017){Soubiran}, {Militzer}, {Driver}, \&
  {Zhang}}]{Soubiran2017}
{Soubiran}, F., {Militzer}, B., {Driver}, K.~P., \& {Zhang}, S. 2017, Physics
  of Plasmas, 24, 041401

\bibitem[{{Steinmetz} {et~al.}(2020){Steinmetz}, {Guiglion}, {McMillan},
  {Matijevi{\v{c}}}, {Enke}, {Kordopatis}, {Zwitter}, {Valentini}, {Chiappini},
  {Casagrande}, {Wojno}, {Anguiano}, {Bienaym{\'e}}, {Bijaoui}, {Binney},
  {Burton}, {Cass}, {de Laverny}, {Fiegert}, {Freeman}, {Fulbright}, {Gibson},
  {Gilmore}, {Grebel}, {Helmi}, {Kunder}, {Munari}, {Navarro}, {Parker},
  {Ruchti}, {Recio-Blanco}, {Reid}, {Seabroke}, {Siviero}, {Siebert}, {Stupar},
  {Watson}, {Williams}, {Wyse}, {Anders}, {Antoja}, {Birko}, {Bland-Hawthorn},
  {Bossini}, {Garc{\'\i}a}, {Carrillo}, {Chaplin}, {Elsworth}, {Famaey},
  {Gerhard}, {Jofre}, {Just}, {Mathur}, {Miglio}, {Minchev}, {Monari},
  {Mosser}, {Ritter}, {Rodrigues}, {Scholz}, {Sharma}, {Sysoliatina}, \& {RAVE
  Collaboration}}]{Rave6}
{Steinmetz}, M., {Guiglion}, G., {McMillan}, P.~J., {et~al.} 2020, \aj, 160, 83

\bibitem[{{Stevenson}(1982)}]{Stevenson1982}
{Stevenson}, D.~J. 1982, \planss, 30, 755

\bibitem[{{Stevenson}(1985)}]{Stevenson1985}
{Stevenson}, D.~J. 1985, \icarus, 62, 4

\bibitem[{{Stevenson} {et~al.}(2022){Stevenson}, {Bodenheimer}, {Lissauer}, \&
  {D'Angelo}}]{Stevenson2022}
{Stevenson}, D.~J., {Bodenheimer}, P., {Lissauer}, J.~J., \& {D'Angelo}, G.
  2022, \psj, 3, 74

\bibitem[{{Sturm} {et~al.}(2022){Sturm}, {McClure}, {Harsono}, {Facchini},
  {Long}, {Kama}, {Bergin}, \& {van Dishoeck}}]{Sturm2022}
{Sturm}, J.~A., {McClure}, M.~K., {Harsono}, D., {et~al.} 2022, \aap, 660, A126

\bibitem[{{Su{\'a}rez-Andr{\'e}s} {et~al.}(2018){Su{\'a}rez-Andr{\'e}s},
  {Israelian}, {Gonz{\'a}lez Hern{\'a}ndez}, {Adibekyan}, {Delgado Mena},
  {Santos}, \& {Sousa}}]{SA2018}
{Su{\'a}rez-Andr{\'e}s}, L., {Israelian}, G., {Gonz{\'a}lez Hern{\'a}ndez},
  J.~I., {et~al.} 2018, \aap, 614, A84

\bibitem[{{Szul{\'a}gyi} {et~al.}(2014){Szul{\'a}gyi}, {Morbidelli}, {Crida},
  \& {Masset}}]{Szul2014}
{Szul{\'a}gyi}, J., {Morbidelli}, A., {Crida}, A., \& {Masset}, F. 2014, \apj,
  782, 65

\bibitem[{{Tabone} {et~al.}(2022){Tabone}, {Rosotti}, {Cridland}, {Armitage},
  \& {Lodato}}]{Tabone2021}
{Tabone}, B., {Rosotti}, G.~P., {Cridland}, A.~J., {Armitage}, P.~J., \&
  {Lodato}, G. 2022, \mnras, 512, 2290

\bibitem[{{Teague} {et~al.}(2019){Teague}, {Bae}, \& {Bergin}}]{Teague2019}
{Teague}, R., {Bae}, J., \& {Bergin}, E.~A. 2019, \nat, 574, 378

\bibitem[{{van der Marel} {et~al.}(2016){van der Marel}, {van Dishoeck},
  {Bruderer}, {Andrews}, {Pontoppidan}, {Herczeg}, {van Kempen}, \&
  {Miotello}}]{vanderMarel2016}
{van der Marel}, N., {van Dishoeck}, E.~F., {Bruderer}, S., {et~al.} 2016,
  \aap, 585, A58

\bibitem[{{van der Walt} {et~al.}(2011){van der Walt}, {Colbert}, \&
  {Varoquaux}}]{numpy}
{van der Walt}, S., {Colbert}, S.~C., \& {Varoquaux}, G. 2011, Computing in
  Science and Engineering, 13, 22

\bibitem[{{van 't Hoff} {et~al.}(2017){van 't Hoff}, {Walsh}, {Kama},
  {Facchini}, \& {van Dishoeck}}]{vtHoff2017}
{van 't Hoff}, M.~L.~R., {Walsh}, C., {Kama}, M., {Facchini}, S., \& {van
  Dishoeck}, E.~F. 2017, \aap, 599, A101

\bibitem[{{Vazan} {et~al.}(2018){Vazan}, {Helled}, \& {Guillot}}]{Vazan2018}
{Vazan}, A., {Helled}, R., \& {Guillot}, T. 2018, \aap, 610, L14

\bibitem[{{Vazan} {et~al.}(2015){Vazan}, {Helled}, {Kovetz}, \&
  {Podolak}}]{Vazan2015}
{Vazan}, A., {Helled}, R., {Kovetz}, A., \& {Podolak}, M. 2015, \apj, 803, 32

\bibitem[{{Vazan} {et~al.}(2016){Vazan}, {Helled}, {Podolak}, \&
  {Kovetz}}]{Vazan2016}
{Vazan}, A., {Helled}, R., {Podolak}, M., \& {Kovetz}, A. 2016, \apj, 829, 118

\bibitem[{{Virtanen} {et~al.}(2020){Virtanen}, {Gommers}, {Burovski},
  {Oliphant}, {Weckesser}, {Cournapeau}, {alexbrc}, {Peterson}, {Reddy},
  {Wilson}, {Haberland}, {Mayorov}, {endolith}, {Nelson}, {van der Walt},
  {Laxalde}, {Brett}, {Polat}, {Larson}, {Millman}, {Lars}, {van Mulbregt},
  {eric-jones}, {Carey}, {Moore}, {Kern}, {Leslie}, {Perktold}, {Striega}, \&
  {Feng}}]{Scipy}
{Virtanen}, P., {Gommers}, R., {Burovski}, E., {et~al.} 2020, {scipy/scipy:
  SciPy 1.5.3}, Zenodo

\bibitem[{{Visser} {et~al.}(2018){Visser}, {Bruderer}, {Cazzoletti},
  {Facchini}, {Heays}, \& {van Dishoeck}}]{Visser2018}
{Visser}, R., {Bruderer}, S., {Cazzoletti}, P., {et~al.} 2018, \aap, 615, A75

\bibitem[{{Wahl} {et~al.}(2017){Wahl}, {Hubbard}, {Militzer}, {Guillot},
  {Miguel}, {Movshovitz}, {Kaspi}, {Helled}, {Reese}, {Galanti}, {Levin},
  {Connerney}, \& {Bolton}}]{Wahl2017}
{Wahl}, S.~M., {Hubbard}, W.~B., {Militzer}, B., {et~al.} 2017, \grl, 44, 4649

\bibitem[{{Wang} {et~al.}(2021){Wang}, {Vigan}, {Lacour}, {Nowak}, {Stolker},
  {De Rosa}, {Ginzburg}, {Gao}, {Abuter}, {Amorim}, {Asensio-Torres},
  {Baub{\"o}ck}, {Benisty}, {Berger}, {Beust}, {Beuzit}, {Blunt}, {Boccaletti},
  {Bohn}, {Bonnefoy}, {Bonnet}, {Brandner}, {Cantalloube}, {Caselli},
  {Charnay}, {Chauvin}, {Choquet}, {Christiaens}, {Cl{\'e}net}, {Coud{\'e} Du
  Foresto}, {Cridland}, {de Zeeuw}, {Dembet}, {Dexter}, {Drescher}, {Duvert},
  {Eckart}, {Eisenhauer}, {Facchini}, {Gao}, {Garcia}, {Garcia Lopez},
  {Gardner}, {Gendron}, {Genzel}, {Gillessen}, {Girard}, {Haubois},
  {Hei{\ss}el}, {Henning}, {Hinkley}, {Hippler}, {Horrobin}, {Houll{\'e}},
  {Hubert}, {Jim{\'e}nez-Rosales}, {Jocou}, {Kammerer}, {Keppler}, {Kervella},
  {Meyer}, {Kreidberg}, {Lagrange}, {Lapeyr{\`e}re}, {Le Bouquin}, {L{\'e}na},
  {Lutz}, {Maire}, {M{\'e}nard}, {M{\'e}rand}, {Molli{\`e}re}, {Monnier},
  {Mouillet}, {M{\"u}ller}, {Nasedkin}, {Ott}, {Otten}, {Paladini}, {Paumard},
  {Perraut}, {Perrin}, {Pfuhl}, {Pueyo}, {Rameau}, {Rodet},
  {Rodr{\'\i}guez-Coira}, {Rousset}, {Scheithauer}, {Shangguan}, {Shimizu},
  {Stadler}, {Straub}, {Straubmeier}, {Sturm}, {Tacconi}, {van Dishoeck},
  {Vincent}, {von Fellenberg}, {Ward-Duong}, {Widmann}, {Wieprecht},
  {Wiezorrek}, {Woillez}, \& {Gravity Collaboration}}]{Wang2021}
{Wang}, J.~J., {Vigan}, A., {Lacour}, S., {et~al.} 2021, \aj, 161, 148

\bibitem[{{Wang} {et~al.}(2020){Wang}, {Leigh}, {Perna}, \& {Shara}}]{Wang2020}
{Wang}, Y.-H., {Leigh}, N. W.~C., {Perna}, R., \& {Shara}, M.~M. 2020, \apj,
  905, 136

\bibitem[{{Wang} {et~al.}(2022){Wang}, {Perna}, {Leigh}, \& {Shara}}]{Wang2022}
{Wang}, Y.-H., {Perna}, R., {Leigh}, N. W.~C., \& {Shara}, M.~M. 2022, \mnras,
  509, 5253

\bibitem[{{Ward}(1991)}]{W91}
{Ward}, W.~R. 1991, in Lunar and Planetary Inst.~Technical Report, Vol.~22,
  Lunar and Planetary Science Conference

\bibitem[{{Wei} {et~al.}(2019){Wei}, {Nomura}, {Lee}, {Ip}, {Walsh}, \&
  {Millar}}]{Wei2019}
{Wei}, C.-E., {Nomura}, H., {Lee}, J.-E., {et~al.} 2019, \apj, 870, 129

\bibitem[{{Wilson} \& {Rood}(1994)}]{WR1994}
{Wilson}, T.~L. \& {Rood}, R. 1994, \araa, 32, 191

\bibitem[{{Woodall} {et~al.}(2007){Woodall}, {Ag{\'u}ndez}, {Markwick-Kemper},
  \& {Millar}}]{Wood07}
{Woodall}, J., {Ag{\'u}ndez}, M., {Markwick-Kemper}, A.~J., \& {Millar}, T.~J.
  2007, \aap, 466, 1197

\bibitem[{{Yelle}(2004)}]{Yelle2004}
{Yelle}, R.~V. 2004, \icarus, 170, 167

\bibitem[{{Zhou} {et~al.}(2022){Zhou}, {Sanghi}, {Bowler}, {Wu}, {Close},
  {Long}, {Ward-Duong}, {Zhu}, {Kraus}, {Follette}, \& {Bae}}]{Zhou2022}
{Zhou}, Y., {Sanghi}, A., {Bowler}, B.~P., {et~al.} 2022, \apjl, 934, L13

\end{thebibliography}

\appendix
\section{Planet formation details} \label{sec:PFdetails}

\subsection{ Gas accretion }

For this work we focus primarily on the overall chemistry of the atmospheres of the PDS 70b and c planets. As a result we ignore the initial build up of the planetary core under the assumption that the core does not contribute significantly to the bulk chemistry in the atmosphere. The gas accretion is limited by a number of different mechanisms depending on the planet's current evolutionary stage. While the planet is still embedded within the protoplanetary disk (ie. has not yet opened a gap) we assume that the gas accretion rate is limited by either the Kelvin Helmholtz timescale (KH) or the Bondi timescale - which ever is longer.

The KH timescale is related to the rate that the collapsing envelope can release its gravitational potential energy as heat. Its timescale has the functional form of \citep{AP18}:\begin{align}
    t_{\rm KH} = 10^7 \rm yr\left(\frac{M_{\rm plnt}}{M_\oplus}\right)^{-2},
    \label{eq:KH_time}
\end{align}
where the exponents are determined by comparing population synthesis models of planetesimal core formation to populations of known exoplanets. In principle the KH timescale depends directly on the opacity of the collapsing envelope, however this is not well constrained. The best fit exponents of Equation \ref{eq:KH_time} include variations over envelope opacity when comparing to the known exoplanet population. 

The Bondi radius describes the region of the disk where the gas is likely to be captured by the planet, the kinetic energy of a gas parcel is less than its gravitational potential energy relative to the planet. The planet can thus not accrete more gas than is available within the Bondi radius, or can be resupplied by viscous processes in the disk. The accretion timescale associated with accreting gas through the Bondi radius is \citep{Dangelo2010}:\begin{align}
    t_B = (C_B\Omega)^{-1} \left(\frac{M_*}{a^2\Sigma}\right)\left(\frac{a}{H}\right)^{-7}\left(\frac{M_{\rm plnt}}{M_*}\right)^{-2},
    \label{eq:Bondi_time}
\end{align}
where $a$ is the orbital radius of the planet, $\Omega$ is the Kepler orbital frequency, and $C_B\simeq 2.6$ is a constant meant to match this simple prescription to full 3D hydrodynamic simulations. The growth of the planet during the embedded phase is thus $\dot{M}_{\rm plnt} = M_{\rm plnt}/{\rm max}(t_{KH},t_B)$.

When the planet becomes sufficiently massive its gravitational influence begins to dominate the surrounding gas over the viscous torques and gas pressure forces, at this point it opens a gap locally in the protoplanetary disk. The criteria for opening a gap in the disk is a planet of mass M$_{\rm plnt}$ such that \citep{Crida2006}:\begin{align}
    \frac{3}{4}\frac{H}{R_H} + \frac{50}{q\mathcal{R}} \lesssim 1,
    \label{eq:gap_open}
\end{align}
where $R_H = a (M_{\rm plnt}/3M_*)^{1/3}$ is the Hill radius of the planet, $q = M_{\rm plnt}/M_*$ is the planet-to-star mass ratio, $\mathcal{R} = a^2\Omega/\nu$ is the Reynolds number, and $\nu = \alpha c_s H$ is the gas viscosity under the standard $\alpha$-prescription of \cite{SS73}.

Once the gap opens the geometry of the accretion flow changes. This is mainly due to the fact that while the gravitational influence on the gas at the disk midplane is strong, straight above the midplane the gravitational force is necessarily weaker. As a result the gas can `leak' across the edge of the gap and lose its pressure support - effectively free falling towards the midplane of the disk, and the planet. This gas motion, often called `meridional flow' \citep[for ex. in ][]{Morbidelli2014}, have been found in high resolution studies of CO gas velocities in HD 163296 to coincide with the expected location of an embedded giant planet \citep{Teague2019}. Furthermore in numerical studies of disk gas hydrodynamics around giant planets, meridional flows have been found to contribute a large fraction \citep[up to $\sim 90\%$,][]{Szul2014} of the gas flux into the planet's region of gravitational influence.

\cite{Morbidelli2014} outlined the rate of gas accretion onto a growing planet that had recently opened a gap. Broadly speaking it is limited by the delivery of material to the outer edge of the gap - the disk accretion rate - however they describe a cycling of material driven by gas falling through a meridional flow into a \textit{de}creting circumplanetary disk and back to the outer edge of the planet-induced gap. This recycled gas returns to hydrostatic equilibrium with the rest of protoplanetary disk gas and can return to the meridional flow and the growing planet. This process evolves on a dynamic timescale rather than the viscous timescale and thus their mass accretion rate into the gap follows \citep{Morbidelli2014}:\begin{align}
    \dot{M}_{\rm gap} = 8\pi\nu\left(r/H\right)\Sigma_{\rm gas}.
    \label{eq:morb_acc}
\end{align}
This accretion rate is faster than the disk equilibrium mass accretion by a factor of $8/3 (r/H)$ which, depending on the disk scale aspect ratio, can be between one and two orders of magnitude higher.

The efficiency that the rate of mass accretion into the gap is transferred to an accretion rate onto the planet depends on the \textit{local} flow around the planet. \cite{Batygin2018} proposed that the magnetic field of the young planet acts to deflect incoming gas into the circumplanetary disk reducing the accretion and growth efficiency. \cite{Cridland2018} derived the connection between the magnetic field strength and the mass accretion efficiency resulting in the scaling:\begin{align}
    \frac{\dot{M}_{\rm plnt}}{M_\oplus/yr} = \frac{4}{3^{3/4}}\left(\frac{R_0}{R_H}\right)^2\left(\frac{M_{\rm plnt}}{M_\oplus}\right)^{-2/7}\left(\frac{\dot{M}_{\rm gap}}{M_\oplus/yr}\right)^{3/7},
    \label{eq:MTG}
\end{align}
where the constant $R_0 = \left(\pi^2/2\mu_0~\mathcal{M}^4/GM^2_\oplus/yr\right)^{1/7}$ has units of length and depends on the magnetic moment $\mathcal{M} = BR^3_{\rm plnt}$ of the (assumed) magnetic dipole of the planet. We assume that the young planet has a (constant) magnetic field strength of 500 Gauss which is two orders of magnitude above that of Jupiter, but less than the typical magnetic field strength of a $\sim$1000 K brown dwarf stars. Interpolating the magnetic field strength in this way is consistent with our general understanding of the geo-dynamo and solar-dynamo \citep{Christensen2009}.

We assume a constant planetary radius of $R_{\rm plnt} = 2 R_{\rm Jupiter}$ during this phase as it is a nominal size of young, self-luminous planets with masses greater than Saturn as reported by \cite{Mordasini15}. \cite{Cridland2018} explored the impact of the planetary size on the final planet mass due to gas accretion through the above mechanism and found that final masses only varied by a factor of a few depending on the whether the planet was a `cold-start' - planetary radius equal to Jupiter's current radius - or a `hot-start' planet with a radius three times larger than Jupiter's current radius.

We employ standard planet formation methods in computing the growth of the synthetic PDS 70 planets. For simplicity, however we keep the disk physical and chemical properties constant throughout the planet formation process. This ignores the fact that the disk generally cools and becomes less dense as a function of time. At later times, as the gas is accreted onto the host star, the surface density of the gas drops and the gas accretion rate onto the planet can be reduced through equation \ref{eq:Bondi_time}. The drop in mass accretion rate will be a particular problem in the viscous model because it already predicts low mass giant planets. The wind model, on the other hand may better predict the masses of the b and c planets if the accretion rate is lower at later times.

\subsection{Planet migration}

Our gas accretion simulations begin with a planetary core mass that is set to the the smaller of the pebble isolation mass \citep{Bitsch2018} and the total dust mass exterior to the initial radii. These masses are a factor of a few lower than the gas gap opening mass \citep{Crida2006} and thus Type I migration will not have a large impact on the overall results of our work. We keep the initial core stationary until it reaches the gas gap opening mass at which point we allow it to migrate via Type-II migration \citep{LP86}. Our choice of ignoring the Type-I phase \citep[ie. ][]{W91} of planet migration is related to our ignoring the chemical impact that the planet core might have on the chemical properties of the atmosphere. As such \textit{where} the core is built is less important to the overall chemical properties of the atmosphere and thus we ignore Type-I migration and the early build up of the core for simplicity.

Once the young planet has reached a mass that satisfies Equation \ref{eq:gap_open} we begin to evolve its orbital radius through standard Type-II migration. In Type-II migration, the planet has opened a gap in the gas which effectively changes the way in which the protoplanetary disk can transport its angular momentum. Because of its connection to the angular momentum transport - typically assumed to be due to viscous evolution in gas - we typically assume that Type-II migration evolves on the viscous timescale:\begin{align}
    \dot{r}_{\rm plnt} = r_{\rm plnt} / t_{\nu},
    \label{eq:type2}
\end{align}
where the viscous time $t_{\nu} = r_{\rm plnt}^2/\nu$ and $\nu = \alpha c_s H$ follows the standard $\alpha$-prescription of \cite{SS73}. The gas viscosity $\nu$ depends on the gas scale height $H$ and the gas sound speed $c_s$\footnote{The DALI model outputs the gas sound speed throughout the disk, for an ideal gas it has the form: $c_s = \sqrt{\gamma R T / \mu {\rm m}_H}$, for an adiabatic constant $\gamma =1$, gas constant $R$, gas temperature $T$, and mean molecular weight of the gas $\mu m_H$}. At a certain point Type-II migration stalls when the planet exceeds a critical mass of M$_{\rm crit} = \pi r^2\Sigma$ - the total mass of the disk gas inward of the planet's current orbital radius. After the planet passes this mass its migration time becomes $t_\nu \rightarrow t_\nu (1 + M_{\rm plnt}/M_{\rm crit})$.

Updated hydrodynamic models concerning Type-II migration have shown that the inner and outer disks are not perfectly separated by the planet-induced gap and some gas crosses the gap \citep{DK15}. The gas crossing the gap can also be accreted by the growing planet \citep{DK17} which reduces the efficiency of the gap crossing. \cite{Robert2018} showed that even given these complications the rate of Type-II migration continues to be proportional to the gas viscosity, which justifies the use of Equation \ref{eq:type2} in the face of more complex hydrodynamic processes. The proportionality is related to the fact that if the planet migrates on a shorter timescale than the viscous timescale, the gas `ahead' of the migrating planet will be compressed by the planetary torques while the space `behind' the planet becomes evacuated. As a result the inner/outer torque would be strengthened/weakened, slowing the migration rate.

\section{Depleting carbon and oxygen abundance}\label{sec:deplete}

\begin{figure}
    \centering
        \includegraphics[width=0.5\textwidth]{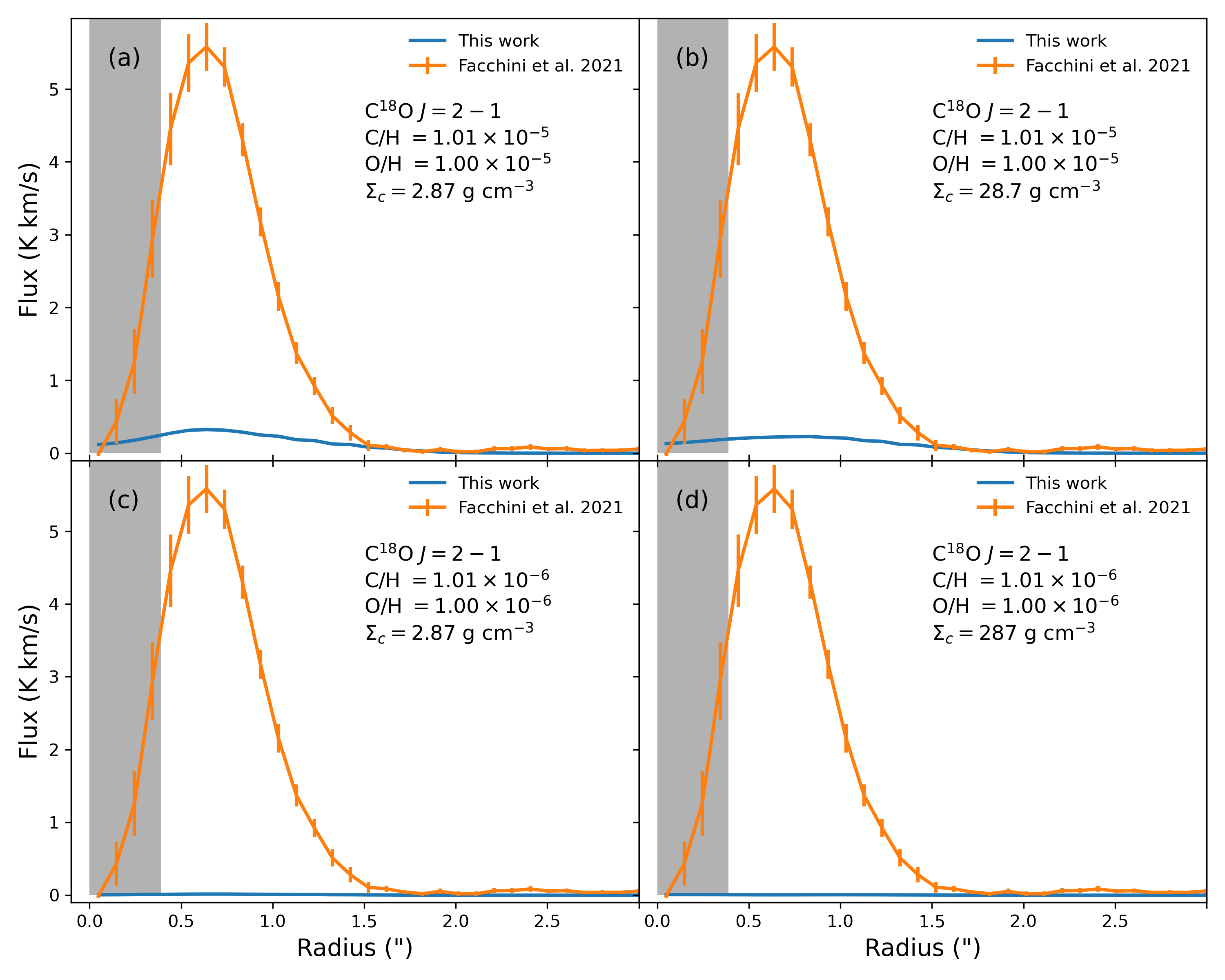}
        \caption{Test of depleting carbon and oxygen abundances in the PDS 70 disk, using C$^{18}$O emission. Each panel represents a different model and their relevant parameters are noted. Panels (a) and (c) use the fiducial critical surface density as in the preferred model in the main text. Panels (b) and (d) have enhanced critical densities by the same factor as the abundances of carbon and oxygen have been depleted. Clearly a depleted abundance is inconsistent with observations.}
        \label{fig:deplete_C18O_quad}
\end{figure}

As mentioned in the main text a common finding in studies focused on C$_2$H flux is that the carbon and oxygen abundances need to be depleted by up to a factor of 100 times the ISM value in order to match to CO isotopologue and H$_2$O observations. The depletion is expected to occur because of the freeze out of CO and H$_2$O onto dust grains that subsequently settle to the midplane - `hiding' the bulk of the volatile abundance of carbon and oxygen. Additional chemical reactions transform CO into other species like CO$_2$, CH$_3$OH, and other hydrocarbons \citep{Bosman2018,Krijt2020}. We have tested how this depletion effects the flux of both C$^{18}$O and C$_2$H in order to better understand our chemical picture.

Figure \ref{fig:deplete_C18O_quad} shows the C$^{18}$O flux from a series of models that are meant to test whether depleting the carbon and oxygen abundances is consistent with observations. In panels (a) and (c) the abundances are depleted, but the density is kept the same as the model presented in Figure \ref{fig:triple_plot}(a). Here we see that we completely lose the C$^{18}$O flux at the edge of the dust ring. In order to possibly recover the flux, we enhance the gas density in the disk by the same factor as the carbon and oxygen abundances are depleted. These tests are shown in panels (b) and (d) and generally show that gas enhancements can not recover the missing flux caused by depleting the carbon and oxygen abundances.

\begin{figure}
    \centering
        \includegraphics[width=0.5\textwidth]{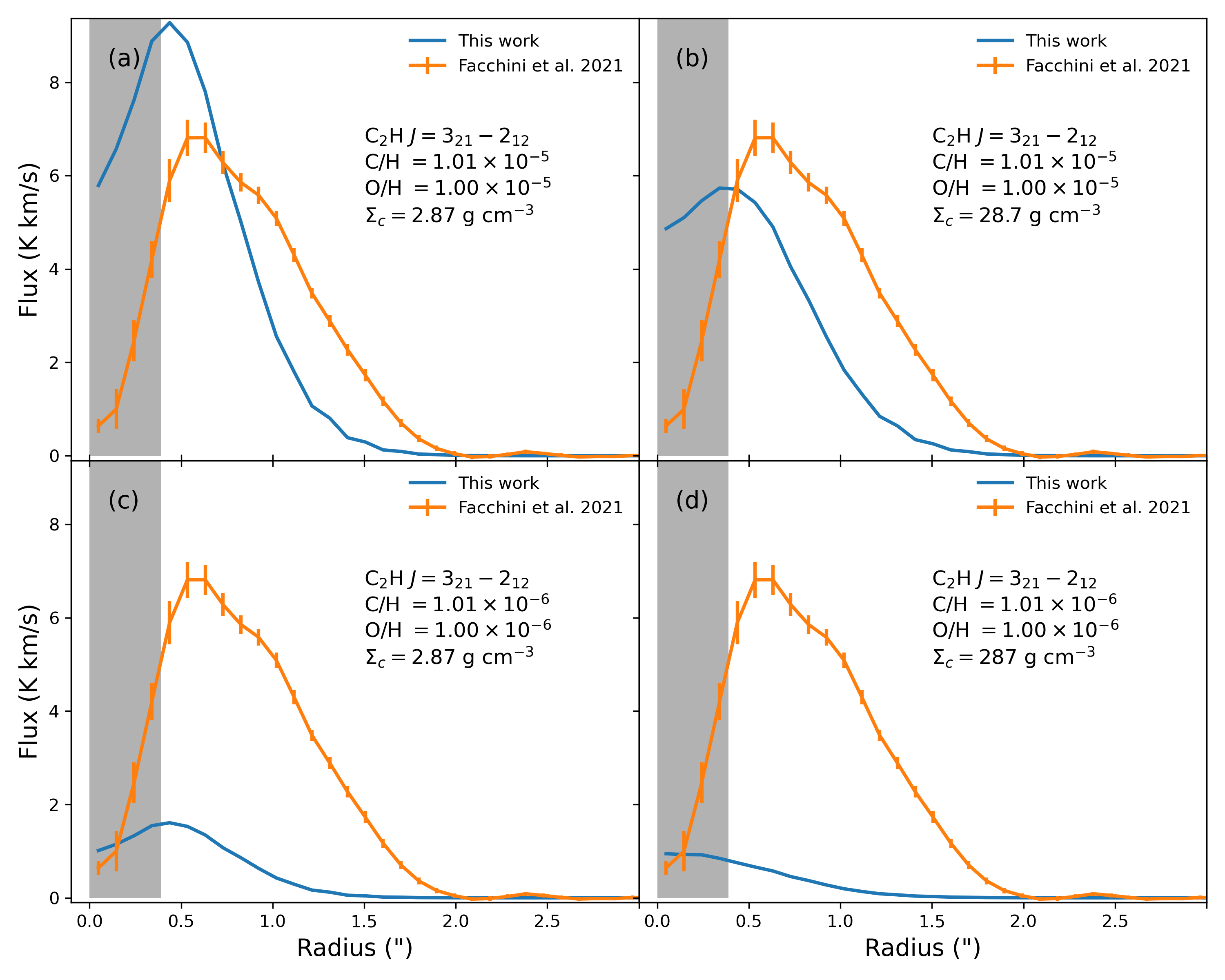}
        \caption{Same as Figure \ref{fig:deplete_C18O_quad} but for C$_2$H emission. Unlike in the C$^{18}$O case, the emission of C$_2$H can remain strong even with depleted abundances. This is consistent with the finding of \cite{Miotello2019}. }
        \label{fig:deplete_C2H_quad}
\end{figure}

In figure \ref{fig:deplete_C2H_quad} we show the C$_2$H flux for same disk models presented in figure \ref{fig:deplete_C18O_quad}. Here we see that the C$_2$H flux is not greatly impacted by a factor of 10 depletion of carbon abundance compared to the fiducial disk model (panel a), and is even weakened when the gas density is enhanced by a factor of 10 (panel b). The reduced flux is related to the equal increase in the dust density in the models where we enhance the critical surface density, since we keep the dust-to-gas ratio constant when we enhance the gas density. The extra dust acts as an extra source of opacity, particularly at the edge of the dust ring, weakening the C$_2$H flux.

The observations are not consistent with a global carbon and oxygen abundances depletion by a factor of 100, which is the depletion done by \cite{Miotello2019}. Here we find a drop in C$_2$H flux by nearly an order of magnitude (panel c), and more so when the gas (and dust) density is enhanced (panel d). Given that we have focused a model with effectively no inner disk - since we use the disk model with $\delta_{\rm gas} = \delta_{\rm dust} = 10^{-15}$ (Figure \ref{fig:deplete_C18O_quad}(a)) we can say that the outer disk is inconsistent in general with a global depletion of carbon and oxygen.

A global depletion of the carbon and oxygen abundance is not consistent with the observed radial profile of \cite{Facchini2021}, however it could help to explain the unusually high C$_2$H flux from the inner disk of our model ($< 6$ AU here). If the inner disk is locally depleted in carbon and oxygen then it would not emit in C$_2$H as brightly as we are predicting in this work. This depletion could be caused by the inefficient transport of material from the outer disk due to the elements being sequestered in the dust and trapped - where observations seem to prefer ISM-like abundances of carbon and oxygen. A future study on the radial distribution of gas and dust, along with their volatile abundances, would be a useful way of furthering our understanding of the PDS 70 system.

\begin{figure}
    \centering
    \includegraphics[width=0.5\textwidth]{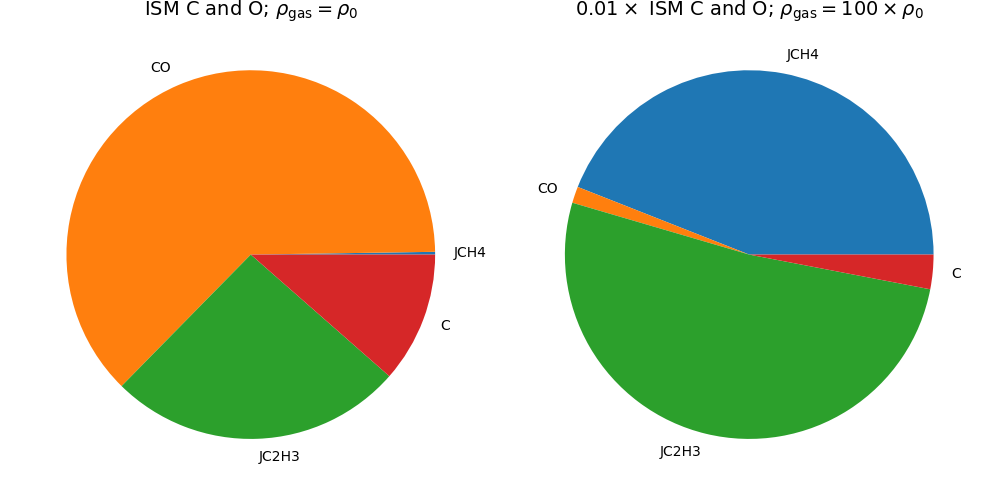}
    \caption{The most abundant species in the model presented in figure \ref{fig:triple_plot}(a) (left panel) and the model presented in figure \ref{fig:deplete_C18O_quad}(d) (right panel), for a disk age of 5 Myr. Molecules with a `J' denotes species that are frozen onto dust grains. The abundances are integrated between 50 AU and 70 AU (ie. near the mm dust ring) and from the midplane to z/r = 0.43. It therefore includes the main reservoir of CO that contributes to the C$^{18}$O flux. It is clear that when the abundances of carbon and oxygen are reduced the main reservoir of carbon is shifted towards hydrocarbons that are frozen onto the dust. }
    \label{fig:most_abundant_spec}
\end{figure}

To understand the sudden drop in C$^{18}$O flux when the elemental abundances of carbon and oxygen are reduced, we show the primary carbon carriers for a pair of models in figure \ref{fig:most_abundant_spec}. The figure shows the relative number of the most abundant carbon carrier in the volatile and ice (labelled with a `J') phases in both models. There we see that when the abundances of carbon and oxygen are reduced by a factor of 100 compared to ISM (right panel) the most dominant carbon carrier in the disk near the dust ring (50-70 AU) is CH$_4$ which remains frozen on the dust grains near the midplane. We understand this shift in the context of the rate of formation of CO, which depends on the number density of both carbon and oxygen. CH$_4$, meanwhile, only depends on the number density of carbon alone and thus when the abundance of carbon and oxygen are reduced, the reduction in production rate that follows is less drastic for CH$_4$ than it is for CO.

\section{Radial profile of $^{13}$CO $J=2-1$}

\begin{figure}
    \centering
    \includegraphics[width=0.5\textwidth]{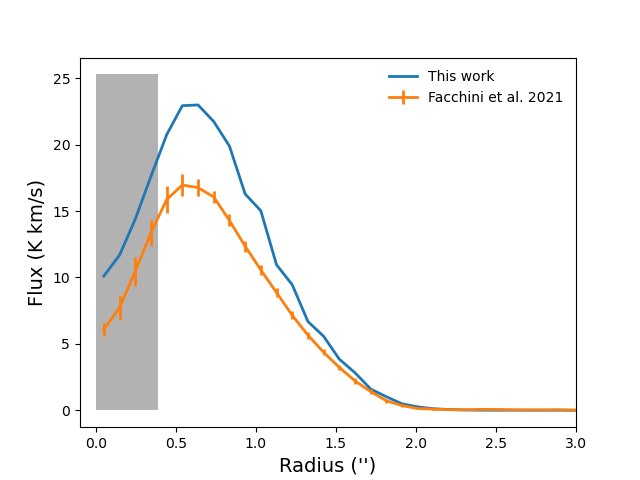}
    \caption{A comparison between the observed $^{13}$CO flux and that computed by the preferred model. $^{13}$CO appears to be partially optically thick as its peak flux is shifted outward relative to the observed flux similarly to the $^{12}$CO flux shown in figure \ref{fig:preferred_plntgap_12CO}.}
    \label{fig:radial_13CO}
\end{figure}

In figures \ref{fig:radial_13CO} we show the radial profiles of the $^{13}$CO emission generated by the preferred model. This line is not used in selecting the preferred model, however we include it here for completeness. We find the peak flux location is slightly shifted outward similarly to $^{12}$CO, while the peak flux is slightly over estimated. This suggests that its emission is slightly optically thick, but not to the same extent as the $^{12}$CO flux.

\end{document}